\documentclass[aps,prd,nofootinbib,twocolumn,superscriptaddress,preprintnumbers,letterpaper,natbib,nofootinbib]{revtex4-1}
\pdfoutput = 1
\usepackage{amssymb}
\usepackage{amsmath}
\usepackage{epsfig}
\usepackage[usenames,dvipsnames]{color}
\usepackage{hyperref}
\usepackage{comment}
\usepackage{color}
\usepackage{cleveref}
\usepackage{multirow}
\usepackage{capt-of}
\usepackage{siunitx}
\usepackage{graphicx}
\usepackage{gensymb}
\usepackage[caption=false]{subfig}
\usepackage{slashed}
\usepackage{tabularx}
\usepackage{enumitem}
\usepackage{multirow}

\hypersetup{
    colorlinks=true,       
    linkcolor=Cyan,          
    citecolor=Magenta}

\definecolor{colorAnn}{rgb}{.2,.7,.2}

\makeatletter
\def\simgt{\mathrel{\lower2.5pt\vbox{\lineskip=0pt\baselineskip=0pt
           \hbox{$>$}\hbox{$\sim$}}}}
\def\simlt{\mathrel{\lower2.5pt\vbox{\lineskip=0pt\baselineskip=0pt
           \hbox{$<$}\hbox{$\sim$}}}}
\makeatother

\newcommand\GBr{\Gamma_{\Phi}^{B}}

\begin{document}

\title{
Baryogenesis and Dark Matter from $\bold{B}$ Mesons
}

\author{Gilly Elor}
\email{gelor@uw.edu}
\affiliation{Department of Physics, Box 1560, University of Washington, Seattle, WA 98195, U.S.A.}

\author{Miguel Escudero}
\email{miguel.escudero@kcl.ac.uk}
\affiliation{From 09/18: Department of Physics, King's College London, Strand, London WC2R 2LS, UK}
\affiliation{Instituto de F\'isica Corpuscular (IFIC), CSIC-Universitat de Val\`encia, Paterna E-46071, Valencia, Spain}

\author{Ann E. Nelson}
\email{aenelson@u.washington.edu}
\affiliation{Department of Physics, Box 1560, University of Washington, Seattle, WA 98195, U.S.A.}

\begin{abstract} 
We present a new mechanism of Baryogenesis and dark matter production in which both the dark matter relic abundance and the baryon asymmetry arise from neutral $B$ meson oscillations and subsequent decays. This set-up is testable at hadron colliders and $B$-factories. 
In the early Universe, decays of a long lived particle produce $B$ mesons and anti-mesons out of thermal equilibrium. These mesons/anti-mesons then 
undergo CP violating oscillations before quickly decaying into visible and dark sector particles. 
Dark matter will be charged under Baryon number so that the visible sector baryon asymmetry is produced without violating the total baryon number of the Universe. 
The produced baryon asymmetry will be directly related to the leptonic charge asymmetry in neutral $B$ decays; an experimental observable.
Dark matter is stabilized by an unbroken discrete symmetry, and proton decay is simply evaded by kinematics. 
We will illustrate this mechanism with a model that is unconstrained by di-nucleon decay, does not require a high reheat temperature, and would have unique experimental signals --   a positive leptonic asymmetry in $B$ meson decays, a new decay of $B$ mesons into a baryon and missing energy, and a new decay of $b$-flavored baryons into mesons and missing energy. These three observables are testable at current and upcoming collider experiments, allowing for a distinct probe of this mechanism.

\end{abstract}

\preprint{KCL-18-53, IFIC-18-35}
\maketitle


\section{Introduction}
\label{sec:Intro}
 
  The Standard Model of Particle Physics (SM), while  now tested to great precision, leaves many questions unanswered. At the forefront of the remaining mysteries is the quest for dark matter (DM); the gravitationally inferred but thus far undetected component of matter which makes up roughly 26\% of the energy budget of the Universe ~\cite{Ade:2015xua,Aghanim:2018eyx}. 
 Many models have been proposed to explain the nature of DM, and various possible production mechanisms to generate the the DM relic abundance -- measured to be $\Omega_{\rm DM} h^2 = 0.1200 \pm 0.0012$~\cite{Aghanim:2018eyx}  -- have been proposed. 
However, experiments searching for DM have yet to shed light on its nature.

Another outstanding question may be stated as follows: why is the Universe filled with complex matter structures when the standard model of cosmology predicts a Universe born with equal parts matter and anti-matter? A dynamical mechanism, {\it Baryogenesis}, is required to generate the primordial matter-antimatter asymmetry;
$Y_B \equiv (n_{B}-n_{\bar{B}})/s = \left( 8.718 \pm 0.004 \right) \times 10^{-11}$, inferred from measurements of the Cosmic Microwave Background (CMB)~\cite{Ade:2015xua,Aghanim:2018eyx} and Big Bang Nucleosynthesis (BBN)~\cite{Cyburt:2015mya,pdg}. A mechanism of Baryogenesis must satisfy the three Sakharov conditions \cite{sakharov};  C and CP Violation (CPV), baryon number violation, and departure from thermal equilibrium.

It is interesting to consider   models and mechanisms that simultaneously generate a baryon asymmetry and produce the DM abundance in the early Universe. For instance, in models of  Asymmetric Dark Matter \cite{Nussinov:1985xr,Dodelson:1989cq,Barr:1990ca,Kaplan:1991ah,Farrar:2005zd,Kaplan:2009ag}, DM carries a conserved charge just as baryons do. Most models of Baryogenesis and/or DM production involve very massive particles and   high  temperatures in the early Universe, making them impossible to test directly and in conflict with cosmologies requiring a low inflation or reheating scale.  

\begin{figure*}[t]
\centering
\includegraphics[width=0.97 \textwidth]{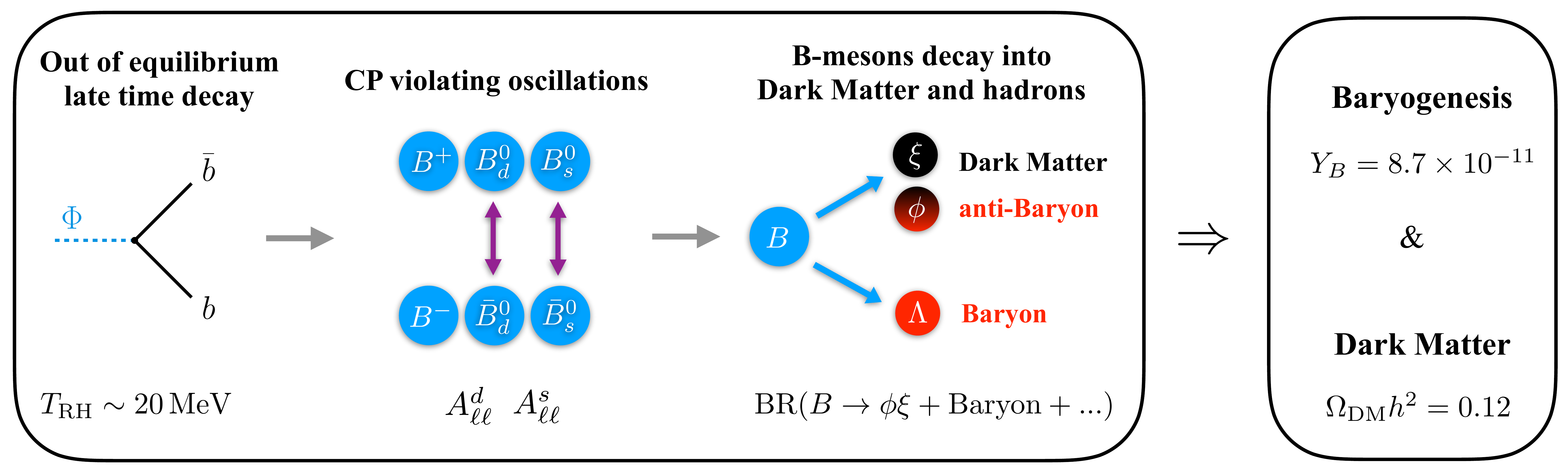}
\caption{Summary of our mechanism for generating the baryon asymmetry and DM relic abundance. $b$-quarks and anti-quarks are produced during a late era in the history of the early Universe, namely $T_{RH} \sim \mathcal{O}(10\,\text{MeV})$, and hadronize into charged and neutral $B$-mesons. The neutral $B^0$ and $\bar{B}^0$ mesons quickly undergo CPV oscillations before decaying out of thermal equilibrium into visible baryons, dark sector scalar baryons $\phi$ and dark Majorana fermions $\xi$. Total Baryon number is conserved and the dark sector therefore carries anti-baryon number. The mechanism requires of a positive leptonic asymmetry in $B$-meson decays ($A_{\ell \ell}^q$), and the existence of a new decay of $B$-mesons into a baryon and missing energy. Both these observables are testable at current and upcoming collider experiments. }
\label{fig:setup}
\end{figure*}

In this work we present a new mechanism for Baryogenesis and DM production that is unconstrained  by nucleon or dinucleon decay, accommodates a low reheating scale $T_{RH} \sim \mathcal{O}(10\,\text{MeV})$, and has distinctive experimental signals.

We will consider a scenario where $b$-quarks and anti-quarks are produced by late, out of thermal equilibrium, decays of some heavy scalar field $\Phi$ (which can be, for instance, the inflaton or a string modulus). The produced quarks hadronize to form neutral $B$-mesons and anti-mesons which quickly undergo CP violating oscillations\footnote{\footnotesize{For instance, the SM box diagrams that mediate the meson anti-meson oscillations contain CP violating phases due to the CKM matrix elements in the quark-$W$ vertices (see for instance \cite{pdg} for a review). Additionally, models of new physics may introduce additional sources of CPV to the $B^0 - \bar{B}^0$ system~\cite{Grossman:1997pa}.}}, 
and decay into a dark sector via a $\Delta B = 0$ four Fermi operator \emph{i.e.} a component of DM is assumed to be charged under baryon number. 
In this way the baryon number violation Sakharov condition is ``relaxed" to an apparent violation of baryon number in the visible sector due to a sharing with the dark sector (in similar spirit to \cite{Agashe:2004bm,Davoudiasl:2010am}). 
The decay of $B$ mesons into baryons, mesons and missing energy would be a distinct signature of our mechanism that can be searched for at experiments such as Belle-II. 
Additionally, the  $\Delta B = 0$ operator allows us to circumvent constraints arising in models with baryon number violation. 

\newpage
We will show that the CPV required for Baryogenesis is  directly related to an experimental observable in neutral $B$ meson decays  -- the leptonic charge asymmetry $A_{\ell \ell}^q$. Schematically, 
\begin{align}
Y_B  \,\, \propto \,\, \sum_{q = s, d} A_{\ell \ell}^q  \times \text{Br}(B_q^0 \to \phi \, \xi + \text{Baryon} + X)\, ,
\end{align}
where we sum over contributions from both $B_s^0 = | \bar{b} \,s \rangle$ and $B_d^0 = |\bar{b} \,d \rangle$, and $ \text{Br}(B_q^0 \to \phi \xi + \text{Baryon} + X) $ is the branching fraction of a $B$ meson into a baryon and DM (plus additional mesons $X$). Note that a positive value of $A_{\ell \ell}^q$ will be required to generate the asymmetry. 
Given a model, the charge asymmetry can be directly computed from the parameters of the $B^0_q$ oscillation system (for instance see \cite{pdg,Artuso:2015swg} for reviews), and as such it is directly related to the CPV in the system.
Meanwhile, $A_{\ell \ell}^q$ is experimentally extracted from a combination of various analysis of LHCb and B-factories  by examining the asymmetry in various $B^0_q$ decays~\cite{pdg}. 

The SM predictions for $A_{\ell \ell}^d$ and $A_{\ell \ell}^s$~\cite{Artuso:2015swg,Lenz:2011ti} are respectively a factor of 5 and 100 smaller than the current constraints on the leptonic asymmetry. Therefore, there is room for new physics to modify $A_{\ell \ell}^{d,\,s}$. 
We will see that since generating the baryon asymmetry in our set-up requires a positive charge asymmetry, there is a region of parameter space where we can get enough CPV from the SM prediction (which is positive) of $A_{\ell \ell}^s$ alone to get $Y_B \sim 10^{-10}$ (provided $A_{\ell\ell}^d = 0$). However, generically the rest of our parameter space will assume new physics.
Note that there are many BSM models that allow for a substantial enlargement of the leptonic asymmetries of both $B_d^0$ and $B_s^0$ systems over the SM values (see \textit{e.g.}~\cite{Artuso:2015swg,Botella:2014qya} and references therein). Note that the flavorful models invoked to explain the recent $B$-anomalies also induce sizable mixing in the $B_s$ system (see \textit{e.g.}~\cite{Altmannshofer:2014cfa,Celis:2015ara,Gripaios:2014tna,Becirevic:2016yqi}).

We summarize the key components of our set-up which will be further elaborated upon in the following sections: 
\begin{itemize}[leftmargin=0.5cm]
\item{A heavy scalar particle $\Phi$ late decays out of thermal equilibrium to $b$ quarks and anti-quarks.}
\item{Since temperatures are low, a large fraction of these $b$ quarks will then hadronize into $B$ mesons and anti-mesons.}
\item{The neutral mesons undergo CP violating oscillations.}
\item{$B$ mesons decay into into the dark sector via an effective $\Delta B = 0$ operator. This is achieved by assuming DM carries baryon number. In this way total baryon number is conserved.}
\item{Dark matter is assumed to be stabilized under a discrete $\mathbb{Z}_2$ symmetry, and proton and dinucleon decay are simply forbidden by kinematics.}
\end{itemize}
Our set up is illustrated in Figure~\ref{fig:setup}, and the details of a model that can generate such a process will be discussed below.
This paper is organized as follows: in Section~\ref{sec:Model} we introduce a model that illustrates our mechanism for Baryogenesis and DM generation, this is accompanied by a discussion of the unique way in which this set-up realizes the Sakharov conditions. Next, in Section~\ref{sec:production} we analyze the visible baryon asymmetry and DM production in the early Universe, by solving a set of Boltzmann equations, while remaining as agnostic as possible about the details of the dark sector. Our main results will be presented here. Next, in Section~\ref{sec:constraints} we discuss the various possible searches that could probe our model, and elaborate upon the collider, direct detection, and cosmological considerations that constrain our model.  
In Section~\ref{sec:DarkDetails} we outline the various possible dark sector dynamics. We conclude in Section~\ref{sec:conclusion}.

\section{Baryogenesis and Dark Matter: Scenario and Ingredients}
\label{sec:Model}

We now elaborate upon the details of our mechanism, and in particular highlight the unique way in which this proposal satisfies the Sakharov conditions for generating a baryon asymmetry. Afterwards we will present the details of an explicit model that will contain all the elements needed to minimally realize our mechanism of Baryogenesis and DM production.

\subsection{Cosmology and Sakharov Conditions}
Key to our mechanism is the late production of $b$-quarks and anti-quarks in the early Universe. To achieve this we assume that a massive, weakly coupled, long lived scalar particle $\Phi$ dominates the energy density of the early Universe after inflation but prior to Big Bang nucleosynthesis. $\Phi$ could be an Inflaton field, a string modulus, or some other particle resulting from preheating. $\Phi$ is assumed to decay, out of thermal equilibrium to $b$-quarks and anti-quarks. 
We only require that $\Phi$ decays late enough so that the Universe is cool enough $\sim \mathcal{O}(10\,\text{MeV})$ for the $b$ quarks to hadronize before they decay \emph{i.e.} 
\begin{align}
T_{\text{BBN}} \,\,\, < \,\,\,  T  \,\,\, <  \,\,\, T_{\text{QCD}} \,.\nonumber
\end{align}
The lower bound ensures that Baryogenesis completes prior to nucleosynthesis.
Note that given a long lived scalar particle late $b$ quark production is rather generic --  there is no obstruction to scenarios in which $\Phi$ decays to   other heavy particles: e.g.  $\Phi$ particles which mainly decays to   $t$-quarks, or Higgs bosons, as these also will promptly decay to $b-quarks$. Furthermore it is very typical for and there is no symmetry preventing scalar particles from mixing with the Higgs Boson and hence primarily decaying into $b$-quarks. For definiteness we will simply assume that $\Phi$ decays out of thermal equilibrium directly into $b$ and $\bar{b}$ quarks. 

The $b$ quarks, injected into the Universe at low temperatures, will mostly hadronize as $B$ mesons -- $B_d^0$, $B_s^0$, and $B^\pm$.  Upon hadronization the neutral $B^0_q$ mesons will quickly undergo CP violating  $B_q^0 - \bar{B}_q^0$ oscillations \cite{pdg}. Such CPV occurs in the SM (and is sizable in the $B$ systems), but  could also  be augmented by new physics. In this way a long lived scalar particle realizes, rather naturally, two of the Sakharov conditions -- departure from thermal equilibrium and CPV.
Interestingly, we will find a region in parameter space where our mechanism can work with just the CPV of the SM, contrary to the usual lore in which the CPV condition must come from beyond the SM physics.

Let us now address the remaining Sakharov condition: baryon number violation. 
While baryon number violation appears in the SM non-perturbatively~\cite{Kuzmin:1985mm}, and is utilized in Leptogenesis models~\cite{Fukugita:1986hr,Dolgov:1991fr,Buchmuller:2005eh,Davidson:2008bu,Drewes:2017zyw}, the SM baryon number violation will be suppressed at the low temperatures we consider here (as it must to ensure the stability of ordinary matter). 
It is possible to engineer models that utilize low scale baryon number violation, but this usually requires an arguably less than elegant construction. For instance, in the setup of \cite{Ghalsasi:2015mxa,Aitken:2017wie} baryon number violation occurred primarily in heavy flavor changing interactions so as to sufficiently suppress the di-nucleon decay rate, which required a very particular flavor structure. 
In the present work, we assume that \emph{DM is charged under baryon number}, thereby allowing for the introduction of new baryon number conserving dark-SM interactions. 

If the $B$ mesons, after oscillations, can quickly decay to DM (plus visible sector baryons), the CPV from  $B^0_q - \bar{B}^0_q$ oscillations will be transferred to the dark sector leading to a matter-antimatter asymmetry in both sectors. 
Critically, the total baryon number of the Universe, which is now shared by both visible and dark sectors, remains zero.  In this way we have ``relaxed" the baryon number violation Sakharov condition to an apparent Baryon number violation in the visible sector. 

\subsection{An Explicit Model}
\label{subsec:Model}
We now present an explicit model which realizes our mechanism. Minimally, we introduce four new particles; a long lived weakly coupled massive scalar particle $\Phi$ (discussed above), an unstable Dirac fermion $\psi$ carrying baryon number, and two stable DM particles -- a Majorana fermion $\xi$ and a scalar baryon $\phi$. 
All are assumed to be singlets under the SM gauge group. To generate effective interactions between the dark and visible sectors, we introduce a TeV mass, colored, electrically charged scalar particle $Y$.
We assume a discrete $\mathbb{Z}_2$ symmetry to stabilize the DM. Table~\ref{table:fields} summarizes the new fields (and their charge assignments) introduced in this model. 
Possible extensions to this minimal scenario will be considered in later sections. 

\subsubsection*{Operators and Charges}

To generate renormalizable interactions between the visible and dark sectors, we a assume a UV model similar to that of \cite{Ghalsasi:2015mxa,Aitken:2017wie}. 
We introduce a $-1/3$ electrically charged, baryon number $-2/3$, color triplet scalar $Y$ which can couple to SM quarks.  Such a new particle is theoretically motivated, for instance $Y$ could be a squark of a theory in which a linear combination of the SM baryon number $U(1)_B$ and a $U(1)_R$ symmetry is conserved \cite{Beauchesne:2017jou}. The details of the exact nature and origin of $Y$ are not important for the present set-up. 
Additionally, we introduce a new neutral Dirac fermion $\psi$ carrying baryon number $-1$. 

\newpage
The renormalizable couplings between $\psi$ and $Y$ allowed by the symmetries include\footnote{\footnotesize{We have suppressed fermion indices for simplicity as there is a unique Lorentz and gauge invariant way to contract fields. In particular, the $s^c$ and $b^c$ are SU(2) singlet right handed Weyl fields. Under $SU(3)_c$, the first term of Equation~\eqref{eq:Lag_psi} is the fully anti-symmetric combination of three $\bar{\bold{3}}$ fields, which is gauge invariant. While the second term is a $\bar{\bold{3}} \times \bold{3} = \bold{1}$ singlet.}}: 
\begin{align}
\label{eq:Lag_psi}
\mathcal{L} \,\,\,\,  \supset \,\,\,\, - \, y_{ub} \,  Y^* \, \bar{u} \, b^c - y_{\psi s} \, Y \bar{\psi} \,  s^c  \,+\, \text{h.c} \, .
\end{align}
We take the mass of the colored scalar to be $m_Y \sim \mathcal{O}(\text{TeV})$ and integrate out the field $Y$ for energies less than its mass, resulting in the following four fermion operator in the effective theory:
\begin{align}
\label{eq:usbOp}
\mathcal{H}_{eff} \,\, = \,\, \frac{y_{ub} y_{\psi s}}{m_Y^2} u \, s\,  b\, \psi \, .
\end{align} 
Other flavor structures may also be present but for simplicity we consider only the effects of the above couplings (see Appendix~\ref{sec:decayoperators} for other possible operators).
Assuming $\psi$ is sufficiently light,  the operator of Equation~\eqref{eq:usbOp} allows the $\bar{b}$-quark within $B_q = | \bar{b} \, q \rangle$ to  decay; $\bar{b} \rightarrow {\psi} \, u \,s$, or equivalently $B_q \rightarrow \psi + \text{Baryon} + X$, where $X$ parametrizes mesons or other additional SM particles.  
Critically, note that $\mathcal{O} = u\, s\, b\,$ in  Equation~\eqref{eq:usbOp} is a $\Delta B = 1$ operator,  so that the operator in Equation~\eqref{eq:usbOp} is baryon number conserving since $\psi$ carries a baryon number $- 1$. 

\begin{table}[t]
\label{Table:trans}
\renewcommand{\arraystretch}{2.0}
\setlength{\arrayrulewidth}{.3mm}
\centering
\small
\setlength{\tabcolsep}{0.36 em}
\setlength{\arrayrulewidth}{.25mm}
\begin{tabular}{ |c || c | c | c | c  | c |}
    \hline
    Field &  Spin &  $Q_{EM}$ &   Baryon no.  & $ \,\, \mathbb{Z}_2 \,\,$  &  Mass \\ \hline \hline
   $\Phi  $ &  $0$  &   $0$ &    $0 $  & $+1$ & $11-100 \, \text{GeV}$ \\ \hline 
   $Y  $ &      $0$  &   $-1/3$ &    $ -2/3$  & $+1$ & $\mathcal{O}(\text{TeV})$ \\ \hline 
   $\psi $ &    $1/2$  &   $0$ &    $-1$  & $+1$ & $\mathcal{O}(\text{GeV})$\\ \hline 
    $\xi  $ &    $1/2$  &   $0$ &    $ 0$  & $-1$& $\mathcal{O}(\text{GeV})$ \\ \hline 
    $\phi$ &   $0$  &   $0$ &     $-1$ & $- 1$ & $\mathcal{O}(\text{GeV})$\\ \hline 
\end{tabular}
\vspace{5mm}
\caption{Summary of the additional fields (both in the UV and effective theory), their charges and properties required in our model.}
\label{table:fields}
\end{table}
\begin{figure}[t]
 \hspace{-0.2cm}
\centering
\includegraphics[width=0.41\textwidth]{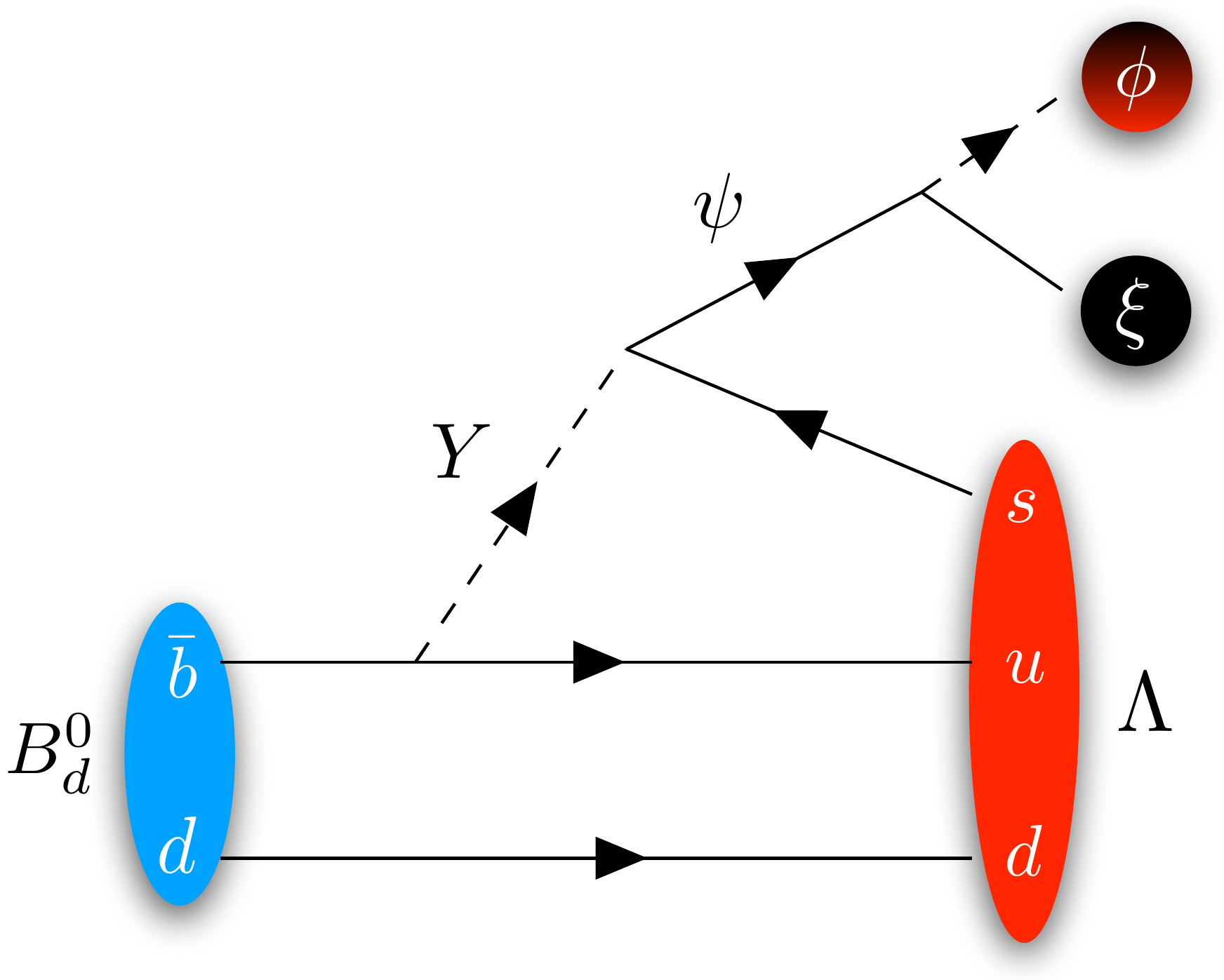}
\caption{An example diagram of the $B$ meson decay process as mediated by the heavy colored scalar $Y$ that results in DM and a visible baryon, through the interactions of Equation~\eqref{eq:Lag_psi} and Equation~\eqref{eq:DarkYukawa}.}
\label{fig:example_decay}
\end{figure}

In this way our model allows for the symmetric out of thermal equilibrium production of $B$ mesons and anti-mesons in the early Universe, which 
subsequently undergo CP violating oscillations \emph{i.e.} the rate for $B^0 \rightarrow \bar{B}^0$ will differ from that of $\bar{B}^0 \rightarrow B^0$. After oscillating the mesons and anti-mesons decay via Equation~\eqref{eq:usbOp} generating an asymmetry in visible baryon/anti-baryon and dark  $\psi$/$\bar{\psi}$ particles (the decays themselves do not introduce additional sources of CPV), so that the total baryon asymmetry of the Universe is zero. 

Since, no net baryon number is produced, this asymmetry could be erased if the $\psi$ particles decay back into visible anti-baryons. Such decays may proceed via a combination of the coupling in Equation~\eqref{eq:usbOp} and weak loop interactions, and are kinematically allowed since $m_\psi > 1.2 \,\text{GeV} $ to ensure the stability of neutron stars \cite{McKeen:2018xwc}.  
To preserve the produced visible/dark baryon asymmetry, the $\psi$ particles should mainly decay into stable DM particles. This is easily achieved by minimally introducing a dark scalar baryon $\phi$ with baryon number $-1$, and a dark Majorana fermion $\xi$. We further assume a discrete $\mathbb{Z}_2$ symmetry under which the dark particles transform as $\psi \rightarrow \psi$, $\phi\rightarrow-\phi$ and $\xi\rightarrow-\xi$. Then the $\psi$ decay can be mediated by a renormalizable Yukawa operator:
\begin{align}
\label{eq:DarkYukawa}
\mathcal{L} \,\, \supset \,\,  -y_d \, \bar{\psi}\, \phi \, \xi\ ,
\end{align}
which is allowed by the symmetries of our model. And in particular, the $\mathbb{Z}_2$ (in combination with kinematic constraints), will make the two dark particles, $\xi$ and $\phi$, stable DM candidates. 

In this way an equal and opposite baryon asymmetry to the visible sector is transferred to the dark sector, while simultaneously generating an abundance of stable DM particles. The fact that our mechanism proceeds through an operator that \emph{conserves} baryon number alleviates the majority of current bounds that would otherwise be very constraining (and would require less than elegant model building tricks to evade). Furthermore, the decay of a $B$-meson (both neutral and charged) into baryons, mesons and missing energy would yield a distinctive signal of our mechanism at B-factories and hadron colliders. An example of a $B$ meson decay process allowed by our model is illustrated in Figure~\ref{fig:example_decay}.

Note that, as in neutrino systems, neutral $B$ meson oscillations will only occur in a coherent system. Additional interactions with the mesons can act to ``measure" the system and decohere the oscillations~\cite{Cirelli:2011ac,Tulin:2012re},  thereby suppressing the CPV and consequently diminishing the generated asymmetry.
Spin-less $B$ mesons do not have a magnetic moment. However, due to their charge distribution, scattering of $e^\pm$ directly off $B$ mesons can still decohere the oscillations (see Appendix~\ref{app:coherence} for details). To avoid decoherece effects, the $B$ mesons must oscillate at a rate similar to or faster than the $e^{\pm} B^0 \rightarrow e^{\pm} B^0$ scattering in the early Universe.

\subsubsection*{Parameter Space and Constraints}

To begin to explore the parameter space of our model we note that the particle masses must be subject to several constraints. 
For the decay $\psi \to \phi \, \xi$ to be kinematically allowed we have the following:
\begin{align}
\label{eq:kinematics}
&m_\phi + m_\xi \,\, < \,\, m_\psi \,.
\end{align}
Note that there is also a kinematic upper bound  on the mass of the $\psi$ such that it is light enough for the decay $B/\bar{B} \rightarrow \psi/\bar{\psi} + \text{Baryon}/\text{anti-Baryon} + \text{Mesons}$ to be allowed. This bound depends on the specific process under consideration and the final state visible sector hadrons produced; for instance in the example of Figure~\ref{fig:example_decay} it must be the case that $m_\psi < m_{B_d^0} - m_{\Lambda} \simeq 4.16\, \text{GeV}$. A comprehensive list of the possible decay processes and the corresponding constraint on the $\psi$ mass are itemized in Appendix~\ref{sec:decayoperators}. 

As mentioned above, DM stability is ensured by the $\mathbb{Z}_2$ symmetry, and the following kinematic condition:
\begin{align}
\label{eq:kinematics}
&| m_\xi - m_\phi | \,\,  < \,\,  m_p + m_e \,.
\end{align}
The mass of a dark particle charged under baryon number must be greater than the chemical potential of a baryon in a stable two solar mass neutron star \cite{McKeen:2018xwc}. This leads to the following bound\footnote{\footnotesize{Note that constraints on bosonic asymmetric DM from the black hole production in neutron stars~\cite{McDermott:2011jp} do not apply to our model. In particular, we can avoid accumulation of $\phi$ particles if they annihilate with a neutron into $\xi$ particles. Additionally, there can be $\phi^4$ repulsive self-couplings which greatly raise the minimum number required to form a black hole.}}:
\begin{align}
\label{eq:neutronstar}
m_\psi  \,\, >  \,\,  m_\phi \,\, > \,\, 1.2 \, \text{GeV} \,.
\end{align}
Additionally, the constraint~\eqref{eq:neutronstar} automatically ensures proton stability. 

The corresponding restrictions on the range of particle masses, along with the rest of our model parameter space, is summarized in Table~\ref{tab:Parameters}. 

Note that since $m_\psi $ must be heavier than the proton, the charmed $D$ meson is too light for our baryogenesis mechanism to work, as $m_D < m_p + m_\psi$ (similarly for the Kaons since $m_K<m_p$). As the top quark decays too quickly to hadronize, the only meson systems in the SM that allow for this  Baryogenesis mechanism are the  neutral $B$ mesons.   

\newpage
\subsubsection*{Dark Sector Considerations}

Throughout this work we remain as model independent as possible regarding additional dark sector dynamics. Our only assumption is the existence of the dark sector particles $\psi$, $\xi$ and $\phi$. In general the dark sector could be much richer; containing a plethora of new particles and forces. Indeed, scenarios in which the DM is secluded in a rich dark sector are well motivated by top-down considerations (see for instance \cite{Essig:2013lka} for a review). Additionally, there are practical reasons to expect (should our mechanism describe reality) a richer dark sector. 

The ratio of DM to baryon energy density has been measured to be $5.36$~\cite{Aghanim:2018eyx}. Therefore, for the case where $\phi$ is the lightest dark sector particle, it must be the case that $m_\phi n_\phi \sim 5 m_p n_B$. 
Since $\xi$ does not carry baryon number and $\psi$ decays completely,  once all of the symmetric $\psi$ component annihilates away we will be left with: $n_B = n_\phi$, implying that $m_\phi \sim 5 m_p$ --  inconsistent with the kinematics of $B$ mesons decays ($m_\phi < m_B-m_\text{Baryon}$).  Introducing additional dark sector baryons can circumvent this problem. 

For instance, imagine adding a stable dark sector state $\mathcal{A}$. We assume $\mathcal{A}$ carries baryon number $Q_{\mathcal{A}}$, and in general be given a charge assignment which allows for $\mathcal{A} - \phi$ interactions (\textit{e.g.} $Q_{\mathcal{A}}=1/3$).  Then the condition that $\rho_{\rm DM} \sim 5 \rho_B$ becomes: $m_\phi n_\phi + m_{\mathcal{A}} n_{\mathcal{A}} \sim 5 m_p n_B $). 
Interactions such as $\phi+ A^*\leftrightarrow \mathcal{A}+ \mathcal{A}$  can then reduce the $\phi$ number density, such that in thermodynamic equilibrium we need only require that $m_{\mathcal{A}} \sim 5 Q_{\mathcal{A}} m_p$, while $\phi$ can be somewhat heavier. In principle ${\mathcal{A}}$ may have fractional baryon number so that both $B$ decay kinematics and proton stability are not threatened. 

Additionally, the visible baryon and anti-baryon products of the $B$ decay are strongly interacting, and as such generically annihilate in the early Universe leaving only a tiny excess of baryons which are asymmetric. 
Meanwhile, the $\xi$ and $\phi$ particles are weakly interacting and have masses in the few GeV range. 
Since, as given the CP violation is at most at the level of $10^{-3}$, the DM will generically be overproduced in the early Universe unless the symmetric component of the DM undergoes additional number density reducing annihilations.
One possible resolution is if the dark sector contained additional states, which interacted with the $\xi$ and $\phi$ allowing for annihilations to deplete the DM abundance so that the sum of the symmetric ($m_\xi n_\xi+m_\phi[n_\phi + n_\phi^\star]$) and the antisymmetric ($m_\phi[n_\phi - n_\phi^\star]$) components match the observed DM density value.

We defer a discussion of specific models leading to the depletion of the symmetric DM component to Section~\ref{sec:DarkDetails}.
In what follows, we simply assume a minimal dark particle content and consider the interplay between $\psi$, $\phi$, and $\xi$ via Equation~\eqref{eq:DarkYukawa}, and account for additional possible dark sector interactions with a free parameter. 

\section{Baryon Asymmetry and Dark Matter Production in the Early Universe}
\label{sec:production}
Using the explicit model of Sec.~\ref{subsec:Model}, we now perform a quantitative computation of the relic baryon number and DM densities. 
We will show that it is indeed possible to produce enough CPV from B meson oscillations to explain the measured baryon asymmetry in the early Universe. Interestingly, there will be a region of parameter space where the positive SM asymmetry in $B_s^0$ oscillations is alone, without requiring new physics contributions, sufficient to generate the matter-antimatter asymmetry. Additionally, we will see that a large parameter space exists that can accommodate the measured DM abundance.   
To study the interplay between  production, decay, annihilation  and radiation in the era of interest we study the corresponding Boltzmann equations. 

\subsection{Boltzmann Equations}
\label{subsec:Boltzmann}
The expected baryon asymmetry and DM abundance  are calculated by solving Boltzmann equations that describe the number and energy density evolution of the relevant particles in the early Universe: the late decaying scalar $\Phi$, the dark particles $\xi$, $\phi$, $\phi^\star$ and radiation ($\gamma, e^\pm, \nu, \, ...$). To properly account for the neutral B meson CPV oscillations in the early Universe one should resort to the density matrix formalism~\cite{Cirelli:2011ac,Tulin:2012re} (which is considerably involved and widely used in Leptogenesis models~\cite{Eijima:2018qke,Drewes:2017zyw}). However, in our scenario, the processes of hadronization, B meson oscillations, decays, as well as possible decoherence effects happen very rapidly ($\tau < \text{ps}$) compared with the $\Phi$ lifetime ($\tau_{\Phi} \sim \text{ms}$). This allows us to work in terms of Boltzmann equations, and we account for possible decoherent effects in a mean field approximation for the B mesons in the thermal plasma. We defer Appendices~\ref{app:coherence} and~\ref{subsec:BannOrdec} to the justifications of the approximations that we use below to simplify the resulting set of Boltzmann equations.

\subsubsection*{Radiation and the $\Phi$ field}
First we describe the evolution of $\Phi$ and its interplay with radiation. 
For simplicity we assume that at  times much earlier than $1/ \Gamma_\Phi$, the energy density of the Universe was dominated by non-relativistic $\Phi$ particles, and  that all of the radiation and matter of the current Universe resulted from $\Phi$ decays. Furthermore, the $\Phi$ decay products are very rapidly converted into radiation, and as such the Hubble parameter during the era of interest is:
\begin{align}
\label{eq:Hubble}
H^2 \equiv \left(\frac{1}{a} \frac{da}{dt}\right)^2  &= \frac{8 \pi}{3 m^2_{Pl}} \left(\rho_{\rm rad} +  m_{\Phi} n_{\Phi} \right) \,.
\end{align}

The Boltzmann equations describing the evolution of the $\Phi$ number density and the radiation energy density read:
\begin{align}
\label{eq:BoltzRad}
  \frac{ d n_{\Phi}}{d t} + 3 H n_\Phi &= - \Gamma_\Phi n_\Phi \, , \\
  \frac{ d \rho_{\rm rad}}{d t} + 4 H \rho_{\rm rad} &=  \Gamma_\Phi   m_\Phi n_\Phi \,,
  \label{eq:rad_den}
  \end{align}
where the source terms on the right-hand side of~\eqref{eq:BoltzRad} describe the $\Phi$ decays which cause the number density of $\Phi$ to decrease as energy is being dumped into radiation. 
Note that if we pick an initial time $t \ll 1/ \Gamma_\Phi$, then $\rho_{\rm rad}$ is small enough that there is no sensitivity to initial conditions and may set $\rho_{\rm rad} =0$.
In practice, we assume that at some high $T>m_{\Phi}$,  $\Phi$ was in
thermal equilibrium with the plasma and that at some temperature $T_\text{dec}$ it decouples; fixing the $\Phi$ number density to $n_{\Phi} \left(T_\text{dec}\right)= \frac{\zeta(3)}{\pi^2} \, T_\text{dec}^3$. This number density serves as our the initial condition and is subsequently evolved using Equation~\eqref{eq:BoltzRad}.  For numeric purposes, we assume that the scalar decouples at $T_\text{dec} = 100 \, \text{GeV}$. We note that, as expected, our results will not be sensitive to the exact decoupling temperature provided $T_\text{dec} > 15 \, \text{GeV}$ \emph{i.e.} when all the SM particles except the top, Higgs and Electroweak bosons are still relativistic.

\subsubsection*{Dark Sector}
The Boltzmann equation for the dark Majorana fermion $\xi$, the main DM component in our model when $m_\xi < m_\phi$, reads:
\begin{align}
\label{eq:Boltzxi}
  \frac{ d n_\xi}{d t} + 3 H n_\xi\,\, =-\langle \sigma v \rangle_\xi \, (n_\xi^2 - n_{\rm eq,\xi}^2) +  2 \,  \GBr   \, n_\Phi \, ,
\end{align}
where we have assumed that the processes of $b$/$\bar{b}$ production, hadronization and decay to the dark sector (see Appendix~\ref{subsec:BannOrdec}), all happen very rapidly on times scales of interest \emph{i.e.}~the $\psi$ particle production and subsequent decay happen rapidly and completely and we need not track the $\psi$ abundance. Therefore, the second term on the right hand side of Equation \eqref{eq:Boltzxi} entirely accounts for the dark particle production via the decays $\Phi \rightarrow B \bar{B} \rightarrow \text{dark sector} + \text{visible}$, and so we have defined:
\begin{align}
\GBr \,\, \equiv \,\, \Gamma_\Phi \,  
 \times  \text{Br}(B \to \phi \xi + \text{Baryon} + X) \,.
\end{align}
Here $ \Gamma_\Phi$  is the $\Phi$ decay width, and  $ \text{Br}(B \to \phi \xi + \text{Baryon} + X)$ is the inclusive branching ratio of $B$ mesons into a baryon plus DM.

The $b$ quarks and anti-quark within all flavors of $B$ mesons and anti-mesons (both neutral and charged $B_{d,s}^0$ and $B^\pm$), will contribute to the $\xi$ abundance via decays through the operators in Equations~\eqref{eq:usbOp} and \eqref{eq:DarkYukawa}. Therefore,  in Equation~\eqref{eq:Boltzxi}, we have  implicitly set the branching fraction of $\Phi$ into charged and neutral $B$ mesons: $\text{Br}({\Phi \to  \bar{B} B}) = 1$.
Note that only the neutral $B^0_{d,s}$ mesons can undergo CP violating oscillations thereby contributing to the matter-antimatter asymmetry. Therefore, we should account for the branching fraction into $B^0_{s,d}$ mesons and anti-mesons when considering the asymmetry.

The first term on the right hand side of Equation~\eqref{eq:Boltzxi} allows for additional interactions, whose presence we require to deplete the symmetric DM component as discussed above.

For the region in parameter space where $m_\xi > m_\phi$, DM is composed of the scalar baryons and anti-baryons, and the DM relic abundance is found by solving for the symmetric component, namely:
\begin{align}
\label{eq:BoltzSym}
\frac{d n_{\phi+ \phi^*}}{d t}  + 3 \, H\, n_{\phi+ \phi^*}  = &- 2\, \GBr    \, n_\Phi  \\   \nonumber
& \,- 2 \, \langle \sigma v \rangle_\phi \left( n_{\phi+\phi^*}^2 - n_{\rm eq,\,\phi+\phi^*}^2 \right) \,.
\end{align}
Analogous to the Boltzmann equation describing the $\xi$ evolution, the second term on the right hand side of Equation~\eqref{eq:BoltzSym} accounts for possible dark sector interactions and self-annihilations, while the first term describes dark particle production via decays. Again we assume the $\psi$ fermion decays instantaneously, and DM can be produced from the decay of both neutral and charged $B$ mesons and anti-mesons.

As previously discussed, DM generically tends to be overproduced in this set-up. Additional interactions are required to deplete the DM abundance in order to reproduce the observed value. 
Whether the DM is composed primarily of $\xi$ or $\phi +\phi^*$, the scattering term in the Boltzmann equations allows for the dark particle abundance to be depleted by annihilations into lighter species. 
In our model, the thermally averaged annihilation cross sections for the fermion and scalar will receive contributions from $\phi-\xi$ generated by the Yukawa coupling of Equation~$ \eqref{eq:DarkYukawa}$ (see Appendix~\ref{app:darkcrosssecs} for rates). This interaction will transform the heavier dark particle population into the lighter DM state.
The annihilation term can, in general, receive contributions from additional interactions.  Therefore, when solving the Boltzmann equations, we simply parametrize additional contributions to $\langle \sigma v \rangle_{\xi}$  and $\langle \sigma v \rangle_{\phi + \phi^*}$  by a free parameter.  In Sec.~\ref{sec:DarkDetails}, we will outline a couple of concrete models that accommodate a depletion of the symmetric DM component.

We have derived Equation~\eqref{eq:BoltzSym} by tracking the particle and anti-particle evolution of the complex $\phi$ scalar using the following Boltzmann equations:
\begin{align}
\label{eq:Boltz_phi}
 &\frac{ d n_{\phi}}{d t} + 3 H n_{\phi} = -\langle \sigma v \rangle_\phi (n_{\phi} n_{\phi^\star} - n_{\rm eq,{\phi}}n_{\rm eq,{\phi^\star}})  \\ 
& \quad \,+\,\GBr \,\, n_\Phi \times \left[ 1 + \sum_q A_{\ell \ell}^q\, \text{Br}({\bar{b} \to  B_q^0}) \, f_{\rm deco}^q  \right]  \,, \nonumber
\end{align}
where we sum over contributions from $B^0_{q = s,d}$ oscillations.
Likewise,
\begin{align}
\label{eq:Boltz_phicomplex}
& \frac{ d n_{\phi^\star}}{d t} + 3 H n_{\phi^\star} = -\langle \sigma v \rangle_\phi (n_{\phi} n_{\phi^\star} - n_{\rm eq,{\phi}}n_{\rm eq,{\phi^\star}})  \\
& \quad \,+\,\GBr \, n_\Phi \times \left[ 1 - \sum_q A_{\ell \ell}^q\, \text{Br}({\bar{b} \to  B_q^0}) \, f_{\rm deco}^q  \right]  \,. \nonumber
\end{align}
Since the the $\phi$ and $\phi^*$ particles are produced via several combinations of meson/anti-meson oscillations and decays, we encapsulate the corresponding decay width difference in a quantity $A_{\ell \ell}^q$ (defined explicitly below in Equation~\eqref{eq:LeptonicAsym}), which is a measure of the CPV in the $B_d^0$ and $B_s^0$ systems. $A_{\ell \ell}^q$ is weighted by a function $f^q_{\text{deco}}$ describing decoherence effects -- these will play a critical role in the evolution of the matter-antimatter asymmetry as we discuss below. For the symmetric DM component, the solution of Equation~\eqref{eq:BoltzSym}, the dependence on $A_{\ell \ell}^q$ cancels off as expected.

Finally, note that Equations~\eqref{eq:BoltzSym} and \eqref{eq:Boltzxi} hold in the regime where the two masses $m_\phi$ and $m_\xi$ are significantly different.
For the case where $m_\phi \sim m_\xi$ coannihilations become important \emph{i.e.} there will be rapid $\phi+\phi^*\leftrightarrow \xi+\xi$ processes mediated by $\psi$ which will enforce a relation between $n_\xi$ and $n_{\phi+\phi^*}$. Specifically, in the non-relativistic limit  $n_\xi/n_\phi=\exp{(m_\phi-m_\xi)/T_D}$, so that the equilibrium abundance depends on the dark sector temperature. It is reasonable to consider a construction where $T_D < |m_\phi - m_\xi |$, so that it is justified to set the equilibrium abundance of the heavier particle to zero. However, since coannihilations represent a very small branch in our parameter space, for simplicity and generality, we simply assume we are far from the regime where coannihilation effects are important so that we can solve Equations~\eqref{eq:Boltzxi},~\eqref{eq:Boltz_phi} and \eqref{eq:Boltz_phicomplex} for the dark sector particle abundances.

\renewcommand{\arraystretch}{1.2}
\begin{table*}[t]
\begin{center}
  \setlength{\arrayrulewidth}{.25mm}
\scalebox{0.93}{
\begin{tabular}{ccccc}
\hline\hline
  Parameter  & Description & Range  & Benchmark Value & Constraint  \\ \hline	
  $m_\Phi$ & $\Phi$ mass & 	  $11-100$ GeV   &	25  GeV	 & -\\
  $\Gamma_\Phi$  & $\Phi$ width  & $ 3\times 10^{-23} < \Gamma_\Phi /\text{GeV} < 5\times 10^{-21}$ & $10^{-22}\, \text{GeV}$ & Decay between $3.5 \, \text{MeV} <T < 30 \, \text{MeV}$ \\
  $m_\psi$ & Dirac fermion mediator & $1.5 \, \text{GeV} < m_\psi < 4.2  \, \text{GeV} $  &  3.3 GeV	& Lower limit from $m_\psi > m_\phi + m_\xi$ \\
  $m_{\xi}$&  Majorana DM  &   $0.3 \, \text{GeV}  < m_{\xi} <  2.7 \, \text{GeV} $   & 1.0 and 1.8 GeV & $|m_\xi- m_\phi| < m_p - m_e$ \\
  $m_{\phi}$&  Scalar DM  &  $ 1.2 \, \text{GeV}  < m_{\phi} <  2.7 \, \text{GeV}$   &	1.5 and 1.3  GeV & $|m_\xi- m_\phi| < m_p - m_e$, $m_\phi > 1.2\,\text{GeV}$\\
   $y_{d}$& Yukawa for $\mathcal{L} = y_d \bar{\psi}\phi \xi$ &    & 0.3  &	$< \sqrt{4\pi}$ \\ \hline \hline
   $ \text{Br}(B \to \phi \xi + \text{..} )$&  Br of $B \to \text{ME} + \text{Baryon}$ & $2\times10^{-4}-0.1$	 & $10^{-3}$	 & $< 0.1$~\cite{pdg} \\
      $A_{\ell \ell}^s$&  Lepton Asymmetry $B_d$ &  $5 \times 10^{-6} <A_{\ell \ell}^d< 8\times10^{-4}$ & 	$6\times10^{-4}$ & $A_{\ell \ell}^d = -0.0021 \pm 0.0017 $~\cite{pdg} \\
   $A_{\ell \ell}^s$&  Lepton Asymmetry $B_s$ &  $10^{-5} <A_{\ell \ell}^s< 4 \times10^{-3}$ & 	$10^{-3}$ & $A_{\ell \ell}^s = -0.0006 \pm 0.0028 $~\cite{pdg} \\ \hline
   $\langle \sigma v \rangle_\phi $& Annihilation Xsec for $\phi$	 & $(6-20) \times 10^{-25} \,\text{cm}^3/\text{s}$     & $10^{-24} \,\text{cm}^3/\text{s}$	&  Depends upon the channel~\cite{Aghanim:2018eyx} \\
   $\langle \sigma v \rangle_\xi $& Annihilation Xsec for $\xi$ &   $(6-20) \times 10^{-25} \,\text{cm}^3/\text{s}$   &	 $10^{-24} \,\text{cm}^3/\text{s}$ & Depends upon the channel~\cite{Aghanim:2018eyx} \\ \hline \hline
\end{tabular} }
\end{center}
\caption{Parameters in the model, their explored range, benchmark values and a summary of constraints. Note that the benchmark values for $ A_{\ell \ell}^q \times \text{Br}(B_q \to \phi \xi + \text{Baryon} + X) $, for $\langle \sigma v \rangle_\phi$ and $\langle \sigma v \rangle_\xi$, are fixed by the requirement of obtaining the observed Baryon asymmetry ($Y_B = 8.7\times 10^{-11}$) and the correct DM abundance ($\Omega_{\rm DM} h^2 = 0.12$), respectively. 
}
\label{tab:Parameters}
\end{table*}

\subsubsection*{Baryon Asymmetry}
The Boltzmann equation governing the production of the baryon asymmetry is simply the difference of the particle and anti-particle scalar baryon abundances Equation~\eqref{eq:Boltz_phi} and Equation~\eqref{eq:Boltz_phicomplex}:
\begin{align}
\label{eq:BaryonAsymmetry}
\nonumber
&  \frac{d (n_\phi - n_{\phi^\star})}{d t} + 3 \, H  (n_\phi- n_{\phi^*})  \\ 
&\quad \quad \quad \quad \quad \,\,=\,\, 2\,\GBr \, \sum_{	q} \text{Br}(\bar{b} \to  B_q^0) \,  A_{\ell \ell}^q \, f_\text{deco}^q \, n_\Phi  \, ,
\end{align}
where we must consider contributions from decays of the $\bar{b}$ anti-quarks/quarks within both $B^0_d$ and $B^0_s$ mesons/anti-mesons: we take the branching fraction for the production of each meson to be $\text{Br}({\bar{b} \to  B^0_d})  = 0.4$ and $\text{Br}({\bar{b} \to  B^0_s})  = 0.1$  according to the latest estimates~\cite{pdg}. 

Interestingly, we see from integrating Equation~\eqref{eq:BaryonAsymmetry} that the baryon asymmetry is fixed by the product $A_{\ell \ell}^q �\times \text{Br}({B_q^0 \to \xi \phi+\text{Baryon} + X})$ -- a measurable quantity at experiments. In particular, $A_{\ell \ell}^q$ is defined as:
\begin{align}
\label{eq:LeptonicAsym}
A_{\ell \ell}^{q} = \frac{\Gamma \left(\bar{B}_q^0 \rightarrow B_q^0 \rightarrow f \right) - \Gamma \left(B_q^0 \rightarrow \bar{B}_q^0 \rightarrow \bar{f}  \right) }{\Gamma \left( \bar{B}_q^0 \rightarrow B_q^0 \rightarrow f \right) + \Gamma \left( B_q^0 \rightarrow \bar{B}_q^0 \rightarrow \bar{f} \right)}\,,
\end{align}
which is directly related to the CPV in oscillating neutral $B$ meson systems.
Here  $f$ and $\bar{f}$ are taken to be final states that are accessible by the decay of $b$/$\bar{b}$ only.  Note that as defined, Equation~\eqref{eq:LeptonicAsym} corresponds to the semi-leptonic asymmetry (denoted by $A_{SL}^q$ in the literature) in which the final state may be tagged. However, at low temperatures and in the limit when decoherence effects are small, this is effectively equivalent to the leptonic charge asymmetry for which one integrates over all times. Therefore, in the present work we will use the two interchangeably.

Maintaining the coherence of $B^0$ oscillation is crucial for generating the asymmetry; additional interactions with the $B$ mesons can act to ``measure" the state of the $B$ meson and decohere the $B^0_q - \bar{B}^0_q$ oscillation \cite{Cirelli:2011ac,Tulin:2012re}, thereby diminishing the CPV and so too the generated baryon asymmetry. 
$B$ mesons, despite being spin-less and charge-less particles, may have sizable interactions with electrons and positrons due to the $B$'s charge distribution. Electron/positron scattering $e^\pm B_q \to e^\pm B_q$, if faster than the $B_q^0$ oscillation, can spoil the coherence of the system. We have explicitly found that this interaction rate is 2 orders of magnitude lower than for a generic baryon~\cite{Aitken:2017wie}, but for temperatures above $T\simeq 20\,\text{MeV}$ the process $\Gamma({e^\pm B \to e^\pm B})$ occurs at a much higher rate than the $B$ meson oscillation and therefore precludes the CP violating oscillation. We refer the reader to Appendix~\ref{app:coherence} for the explicit calculation of the $e^\pm B \to e^\pm B$ scattering process in the early Universe.

Generically, decoherence will be insignificant if oscillations  occur at a rate similar or faster then the $B^0$ meson interaction. 
By comparing the $e^\pm B_q \to e^\pm B_q$ rate with the oscillation length $\Delta m_{B_q}$, we construct a step-like  function (we have explicitly checked that a Heaviside function yields similar results) to model the loss of coherence of the oscillation system in the thermal plasma:
\begin{align}\label{eq:dec_f}
f_{\rm deco}^q=e^{-\Gamma \left(e^\pm B_q^0 \to e^\pm B_q^0\right) /\Delta m_{B_{q}}} \,.
\end{align}
We take $\Delta m_{B_d} = 3.337\times 10^{-13} \,\text{GeV}$ and $\Delta m_{B_s} = 1.169\times 10^{-11} \,\text{GeV}$~\cite{pdg}, and $\Gamma \left(e^\pm B_q^0 \to e^\pm B_q^0\right) = 10^{-11} \, \text{GeV}\, ( {T}/{20 \, \text{MeV} })^5$ (see Appendix~\ref{app:coherence} for details).

Even without numerically solving the Boltzmann equations, we can understand the need for additional interactions in the dark sector $\langle \sigma v \rangle_{\xi, \phi}$. 
From Equations~\eqref{eq:Boltzxi} and \eqref{eq:BoltzSym}, we see that the DM abundance is sourced by $\text{Br}(B \to \phi \xi + \text{Baryon} + X))$; the greater the value of this branching fraction, the more DM is generated. 
From Equation~\eqref{eq:BaryonAsymmetry}, we see that the asymmetry also depends on this parameter but weighted by a small number; $A^q_{\ell \ell} < 4\times 10^{-3}$. Therefore, generically a region of parameter space that produces the observed baryon asymmetry will overproduce DM, and we require additional interactions with the DM to deplete this symmetric component and reproduce $\Omega_{\rm DM} h^2 = 0.120$. 

\subsection{Numerics and Parameters} 

\begin{figure*}[t]
\begin{tabular}{cc}
		\label{fig:Y_Baryo1}
		\includegraphics[width=0.48\textwidth]{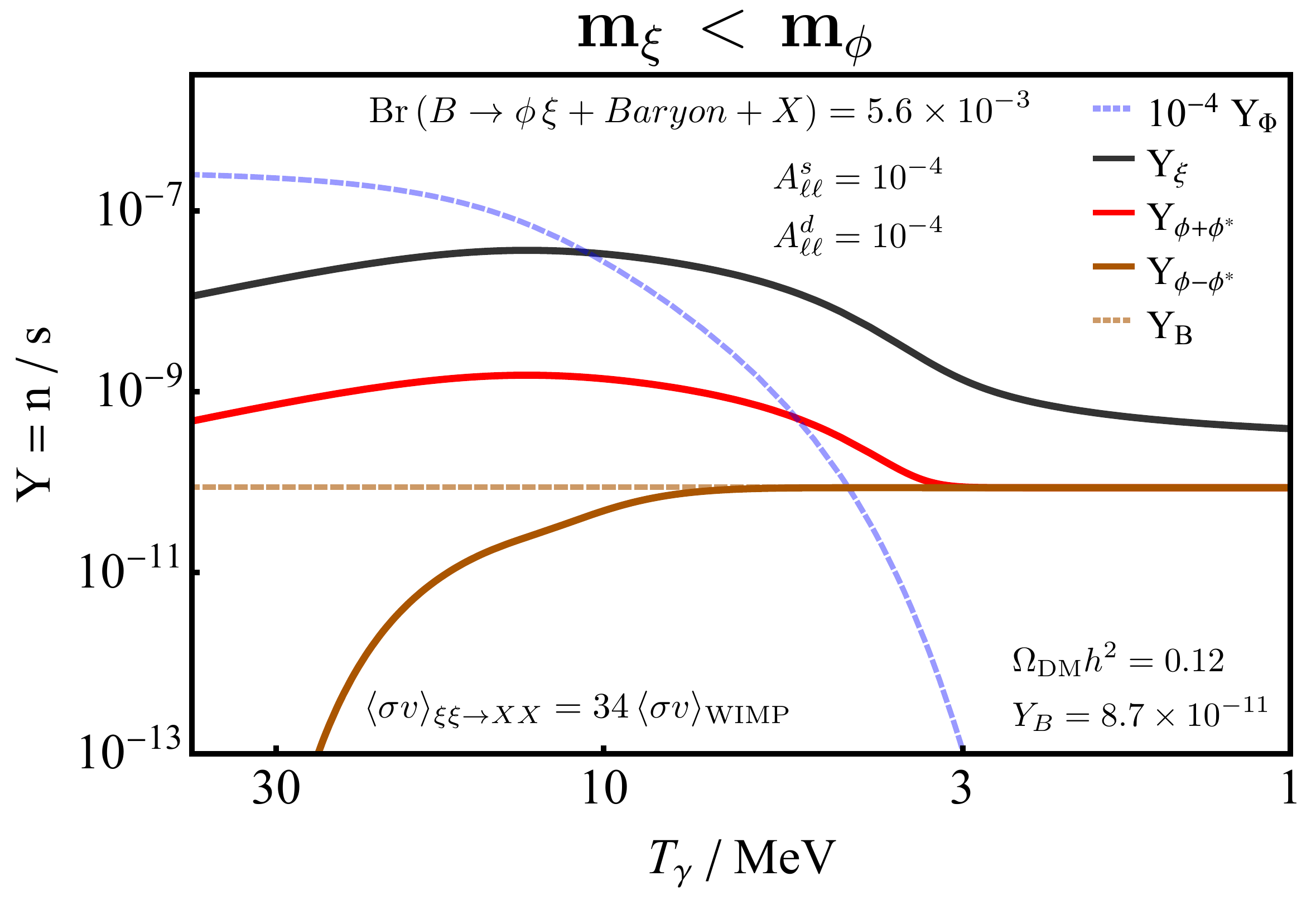}
&
		\label{fig:Y_Baryo2}
		\includegraphics[width=0.48\textwidth]{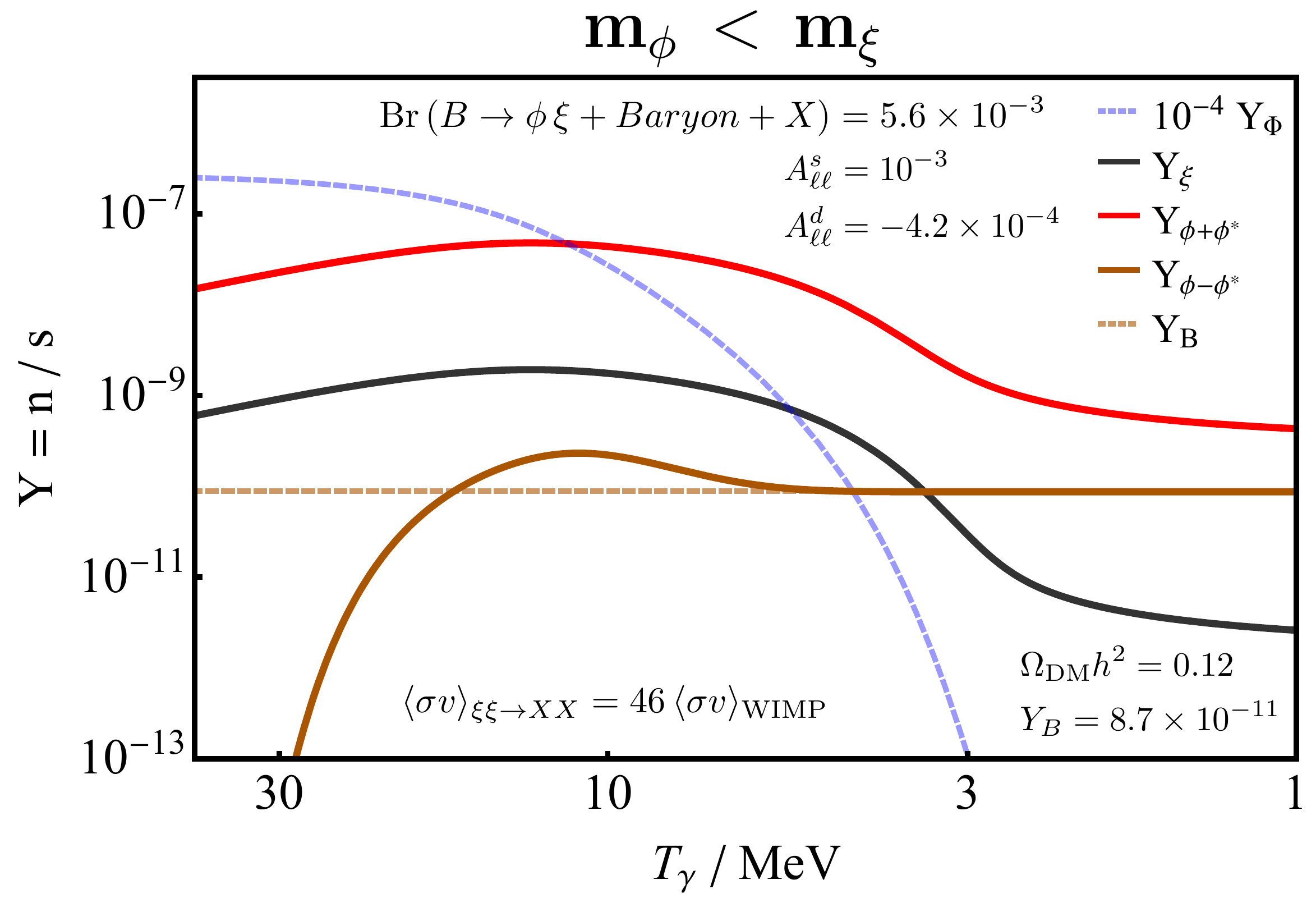}
\end{tabular}
  \caption{Evolution of comoving number density of various components for the benchmark points we consider in Table~\ref{tab:Parameters}: 
  $\{m_\Phi ,\, \Gamma_\Phi ,\, \text{Br}({B  \to \xi \phi}+\text{Baryon}) ,\, m_\Psi ,\,  y_d  \} = \{25 \, \text{GeV},\, 10^{-22}\,\text{GeV},\, 5.6 \times 10^{-3} ,\, 3.3\,\text{GeV},\, 0.3\}$. The \textit{left panel} corresponds to the DM mainly composed of Majorana $\xi$ particles, as we take $m_\xi = 1\,\text{GeV}$ and $m_\phi = 1.5\,\text{GeV}$. We take both the $B_s^0$ and the $B_d^0$ contributions to the leptonic asymmetry to be positive, $A_{\ell \ell}^s = 10^{-4} = A_{\ell \ell}^d $. The change in behavior of the asymmetric yield at  $T \sim 15\,\text{MeV}$ corresponds to decoherence effects spoiling the $B_d^0$ oscillations while $B_s^0$ oscillations are still active.
  The \textit{right panel} corresponds to the DM being composed mainly of dark baryons $\phi+ \phi^*$, with $m_\phi = 1.3\,\text{GeV}$ and 
  $m_\xi = 1.8\,\text{GeV}$. We now take $A_{\ell \ell}^s = 10^{-3}$, and $A_{\ell \ell}^d=A_{\ell \ell}^d |^{\rm SM} = -4.2\times 10^{-4}$ -- the dip in the asymmetry can be understood from the negative value of $A_{\ell \ell}^d$ chosen in this case to correspond to the SM prediction. Both benchmark points reproduce the observed DM abundance $\Omega_{\rm DM} h^2 =0.12$, and baryon asymmetry $Y_B = 8.7 \times 10^{-11}$. 
  }
  \label{fig:Y_Baryo}
\end{figure*}

We use Mathematica \cite{Mathematica} to numerically integrate the set of Boltzmann Equations ~\eqref{eq:BoltzRad},~\eqref{eq:rad_den},~\eqref{eq:Boltzxi},~\eqref{eq:BoltzSym}, and~\eqref{eq:BaryonAsymmetry} subject to the constraint Equation~\eqref{eq:Hubble}. To simplify the numerics it is useful to use the temperature $T$ as the evolution variable instead of time. Conservation of energy yields the following relation~\cite{Scherrer:1987rr,Hannestad:2004px}:
\begin{align}
\frac{dT}{dt} = - \frac{3 H (\rho_{\rm SM} + p_{\rm SM}) - \Gamma_\Phi n_\Phi m_\phi }{d \rho_{\rm SM}/dT} \, ,
\end{align}
which above the neutrino decoupling temperatures $T \gtrsim 3 \, \text{MeV}$ simplifies to~\cite{Venumadhav:2015pla}:
\begin{align}
\label{eq:dTdt}
\frac{dT}{dt} = - \frac{4 H g_{*,s} T^4 -\frac{30}{\pi^2} \times \Gamma_\Phi m_\Phi n_\Phi}{T^3\left[4 \, g_*  + T\, {dg_*}/{dT}\right]} \,.
\end{align}
We can therefore use Equation~\eqref{eq:dTdt} in place of Equation~\eqref{eq:rad_den}. 
For the number of relativistic species contributing to entropy and energy $g_{*,s}(T)$ and $g_*(T)$, we use the values obtained in~\cite{Laine:2006cp}. Finally, since the DM particles generically have masses greater than a $\text{GeV}$ we can safely neglect the inverse scatterings in the DM Boltzmann equations \emph{i.e.} the $n_{eq}^2$ term. To make the integration numerically straightforward we change variables and solve the equations for $\log n$ and $\log T$, such that $\frac{d \log n}{d \log T} = \frac{T}{n} \frac{d n}{dT}$. Note, that we also convert to the convenient yield variables $Y_x = n_x/s$. 

The parameter space of our model includes the particle masses, the $\Phi$ decay width, the dark Yukawa coupling, the branching ratio of $B$ mesons to DM and hadrons, the leptonic asymmetry, and the dark sector annihilation cross sections. Table~\ref{tab:Parameters} summarizes the parameters and the range of over which they are allowed to vary taking into account all constraints. 

 The upper limit on the $\Phi$ mass is imposed because above $\sim 100 \,\text{GeV}$, the scalar could potentially have a small  branching fraction to $b$ quarks (see \emph{e.g.} \cite{Djouadi:1997yw}).

DM masses are constrained by kinematics, and neutron star stability -- Equations~\eqref{eq:kinematics} and \eqref{eq:neutronstar}. 
We take the Yukawa coupling in the dark sector to be $0.3$ since this value enables an efficient depletion of the heavier DM state to the lower one, thus simplifying the phenomenology. For sufficiently lower values of this coupling we may require interactions of both the $\xi$ and $\phi$ states with additional particles. 
The current bounds~\cite{pdg} on the leptonic asymmetry read $A_{\ell \ell}^d = -0.0021 \pm 0.0017$ and $A_{\ell \ell}^s = -0.0006 \pm 0.0028 $ for the $B_{d}^0$ and $B_s^0$ systems respectively.  
Note that these values allow for additional new physics contributions beyond those expected from the SM alone~\cite{Artuso:2015swg,Lenz:2011ti}: $A^s_{\ell \ell}|_{\textrm{SM}}  = (2.22\pm 0.27) \times 10^{-5}$ and $A^d_{SL}|_{\textrm{SM}}= (-4.7\pm 0.6)\times 10^{-4}$.
While there is no direct search for the branching ratio $\text{Br}({B \to \xi \phi+\text{Baryon} + X})$, we can constrain the range of experimentally viable values. For instance, in the example of Figure~\ref{fig:example_decay} where the produced baryon is a $\Lambda = | u \, d\, s \rangle$, we can, based on the $B^+$ decay to $cX$, set the bound $\text{Br}({B  \to \xi \phi+\text{Baryon}}) < 0.1$ at 95\% CL~\cite{pdg}.

\subsection{Results and Discussion}\label{sec:results}
\begin{figure*}[t!]
\centering
\begin{tabular}{cc}
 \includegraphics[width=0.51\textwidth]{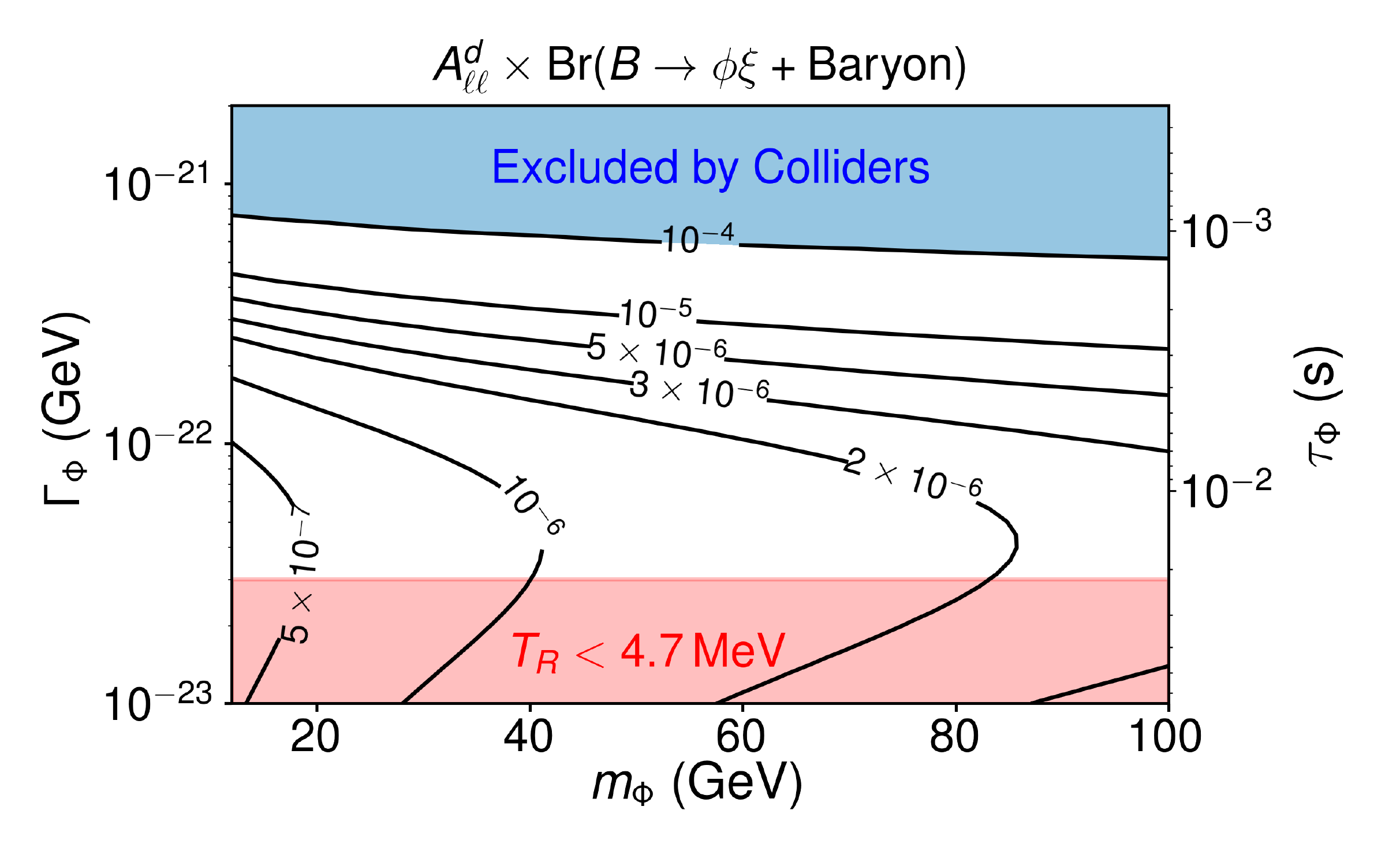} &
  \includegraphics[width=0.51\textwidth]{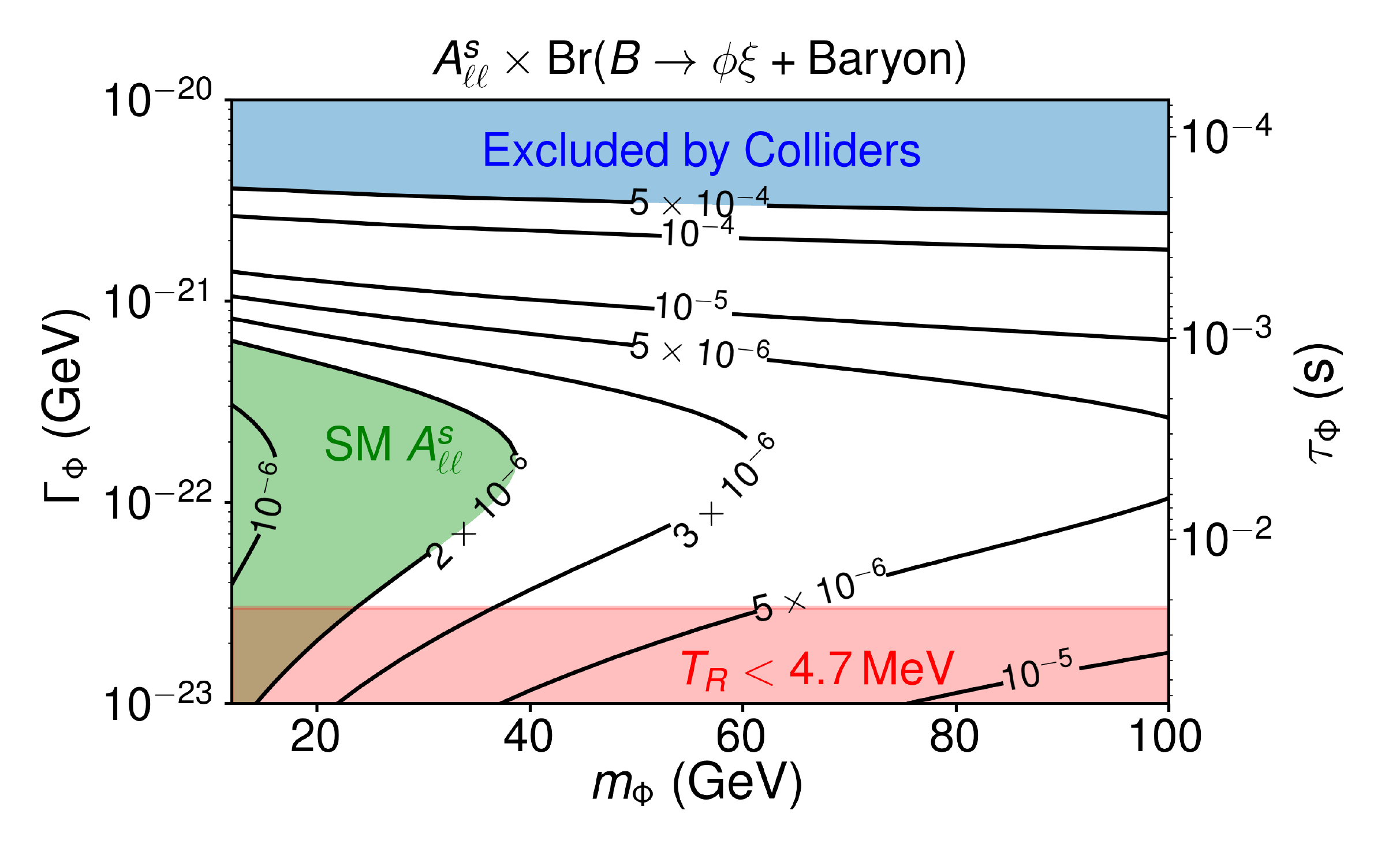}
  \end{tabular}
\caption{\textit{Left panel:} required value of $A_{\ell \ell}^d \times \text{Br}(B\to\xi\phi +\text{Baryon})$ assuming $A_{\ell \ell}^s = 0$ to obtain $Y_B = 8.7\times 10^{-11}$. \textit{Right panel:} Required value of $A_{\ell \ell}^s \times \text{Br}(B\to\xi\phi +\text{Baryon})$ assuming $A_{\ell \ell}^d = 0$ to obtain $Y_B = 8.7\times 10^{-11}$. The blue region is excluded by a combination of constraints on the leptonic asymmetry and the branching ratio~\cite{pdg}. The lower bound (red region) comes from requiring the late $\Phi$ decays to not  spoil the measured effective number of neutrino species from CMB and the measured primordial nuclei abundances~\cite{deSalas:2015glj}.}
\label{fig:BRAll}
\end{figure*}

The recent Planck CMB observations imply a comoving baryon asymmetry of $Y_B = (n_{B}-n_{\bar{B}})/s = \left( 8.718 \pm 0.004 \right) \times 10^{-11}$~\cite{Aghanim:2018eyx}. 
In our scenario, even without fully solving the system of Boltzmann equations,  we can see from integrating Equation~\eqref{eq:BaryonAsymmetry} that the baryon asymmetry directly depends upon the product of leptonic asymmetry times branching fraction:
\begin{align}
Y_B  \,\, \propto \,\, \sum_{q = s, d} A_{\ell \ell}^q  \times \text{Br}(B_q^0 \to \phi \xi + \text{Baryon} + X)\, .\nonumber
\end{align}
Meanwhile, the DM relic abundance is measured to be $\Omega_{\rm DM} h^2 = 0.1200 \pm 0.0012$~\cite{Aghanim:2018eyx} and reads $\Omega_{\rm DM} h^2 = \left[ m_\xi Y_\xi + m_\phi(Y_\phi + Y_{\phi^\star})\right] s_0 h^2/\rho_c$ (where $s_0$ is the current entropy density and $\rho_c$ is the critical density).
In Figure~\ref{fig:Y_Baryo} we display the results (the comoving number density of the various components) of numerically solving the Boltzmann equations for two sample benchmark points that reproduce the observed DM abundance and baryon asymmetry.

Consider the plot on the right panel of Figure~\ref{fig:Y_Baryo}, which corresponds to the case where DM is composed of $\phi$ and $\phi^*$ particles. We can understand the behavior of the particle yields as follows: $\Phi$ particles start to decay at $T \sim 50 \, \text{MeV}$, thereby increasing the abundance of the dark particles $\xi$ and $\phi+\phi^*$ until $T \sim 10 \, \text{MeV}$ at which point $\Phi$ decay completes (as it must, so that the predictions of BBN are preserved). The dip in the dark particle yields at lower temperatures is the necessary effect of the additional annihilations -- which reduce the yield to reproduce to the observed DM abundance. Meanwhile, the asymmetric component $Y_{\phi} - Y_{\phi^*}$ is only generated for $T\lesssim 30\,\text{MeV}$, as it is only then that the $B_s^0$ CPV oscillations are active in the early Universe. 
The decrease in the asymmetric component at $T\sim 10\,\text{MeV}$ is due to the negative contribution of the $B_d^0$ decays, since in this case the leptonic asymmetry is chosen to be negative. Note that for the case in which the DM is mostly composed of $\phi$ and $\phi^*$ particles the observed baryon asymmetry and DM abundance imply an asymmetry of 
\begin{align}
\frac{Y_\phi-Y_\phi^*}{Y_\phi+Y_\phi^*} =  \frac{\Omega_b h^2}{\Omega_{\rm DM}h^2 } \frac{m_\phi}{m_p} \simeq \frac{1}{5.36} \frac{m_\phi}{m_p} \,.
\end{align}

The plot in the left panel of Figure~\ref{fig:Y_Baryo} corresponds to the case where DM is mostly comprised of $\xi$ particles. In this case the evolution of the dark particles is rather similar. Here we have chosen $A_{\ell \ell}^d = A_{\ell \ell}^s > 0$, so that the asymmetric component gets two positive contributions at $T \lesssim  30\,\text{MeV}$ from both $B_d^0$ and $B_s^0$ CPV oscillations. While at $T \sim 15\,\text{MeV}$ the change in behavior of the yield curve corresponds to the contribution from the $B_d^0$ oscillations --  given that the $B_s$ oscillation time scale is 20 times smaller than the $B_d$ one, and the $B_s$ contribution it is active at higher temperatures.

\subsubsection*{The Baryon Asymmetry}

In order to make quantitative statements, beyond the benchmark examples discussed above, we have explored the parameter space outlined in Table~\ref{tab:Parameters} and mapped out the regions that reproduce the observed baryon asymmetry of the Universe. 
From Equation~\eqref{eq:BaryonAsymmetry}, we see the baryon asymmetry depends on the product of the leptonic asymmetry times branching fraction (with contributions from both $B_d^0$ and $B_s^0$ mesons), as well as the $\Phi$ mass and width. The result of this interplay is displayed in Figure~\ref{fig:Y_Baryo}, where the contours correspond to the value the product of $A_{\ell \ell}^q  \times \text{Br}(B^0_s \to \phi \xi + \text{Baryon}+X)$ needed to reproduce the asymmetry $Y_B = 8.7 \times 10^{-11}$ for a given point in $\left(m_\Phi, \Gamma_\Phi \right)$ space. For simplicity, the left and right panels show the effects of considering either the $B^0_d$ or the $B^0_s$ contributions but generically both will contribute.

While the entire parameter space in Figure~\ref{fig:BRAll} is allowed by the range of uncertainty in the experimentally measured values of $A_{\ell \ell}^q$, our range of prediction is further constrained. In particular, the blue region in Figure~\ref{fig:BRAll} is excluded by a combination of constraints on the leptonic asymmetry and the branching ratio~\cite{pdg} (see Section~\ref{sec:constraints}), while the lower bound comes from requiring that the $\Phi$ not spoil the measured effective number of neutrino species from CMB and the measured primordial nuclei abundances~\cite{deSalas:2015glj}.
Therefore, to reproduce the expected asymmetry coming from, for instance, only $B_s^0$, we find $A_{\ell \ell}^s  \times \text{Br}(B \to \phi \xi + \text{Baryon}+X) \sim 10^{-6}- 5 \times 10^{-4}$ (depending upon the $\Phi$ width and mass). 

\begin{figure}[t]
  \includegraphics[width=0.51 \textwidth]{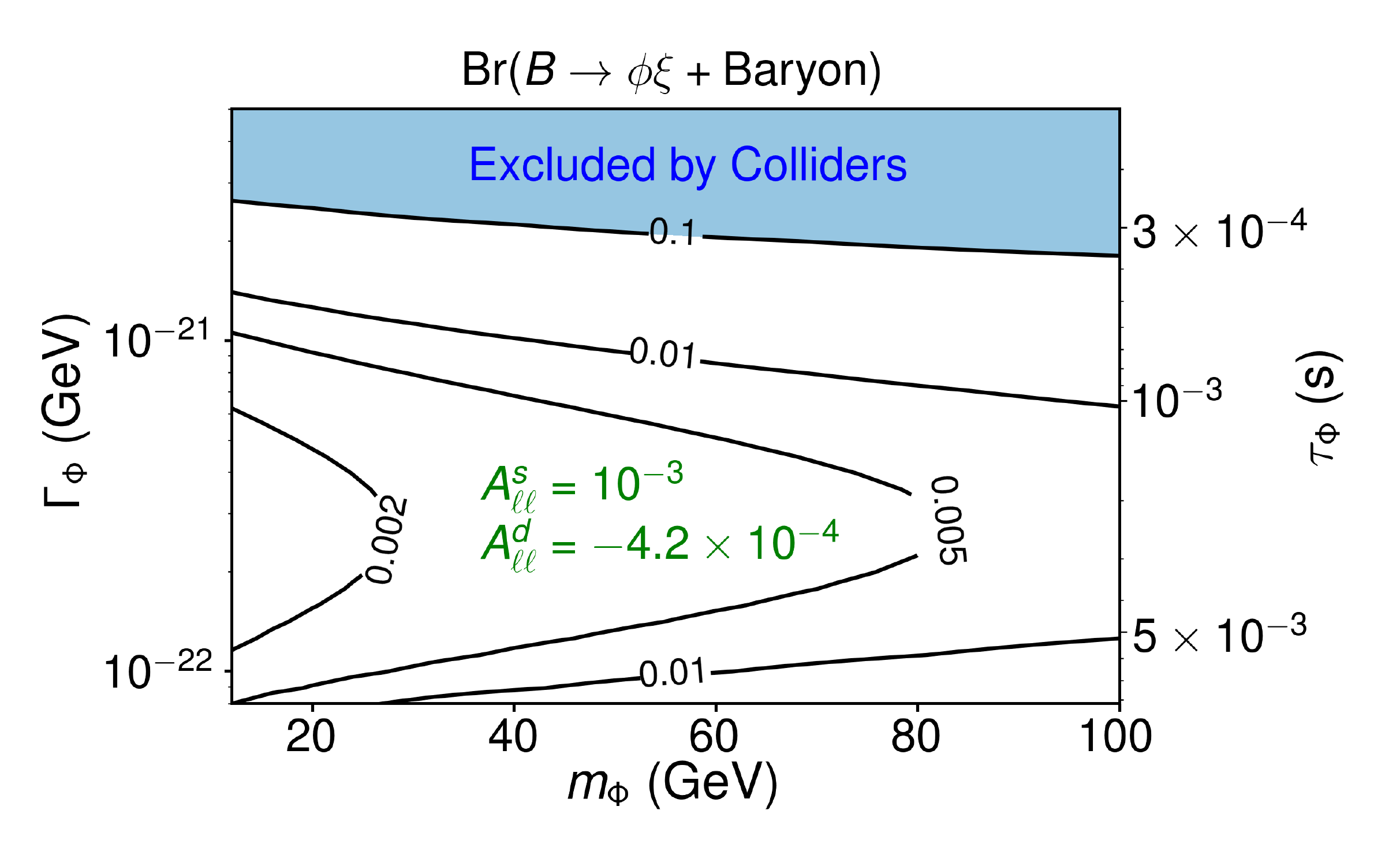}
\caption{Contours show the value $\text{Br}(B\to\xi\phi +\text{Baryon})$ required to generate the correct Baryon asymmetry $Y_B = 8.7\times 10^{-11}$ for the fixed values: $A_{\ell \ell}^s = 10^{-3}$ and $A_{\ell \ell}^d = A_{\ell \ell}^d |^{\rm SM} = -4.2\times 10^{-4}$.}
\label{fig:BR_fixed}
\end{figure}

Interestingly, the baryon asymmetry can be generated with only the SM leptonic asymmetry $A_{\ell \ell}^s = 2\times 10^{-5}$, provided that $\text{Br}(B \to \phi \xi + \text{Baryon}+X) = 0.05-0.1$ and that $A_{\ell \ell}^d = 0$ (which is compatible with current data)  -- see the green region in the right panel of Figure~\ref{fig:BRAll}. Additionally, if new physics enhances $A_{\ell \ell}^s$ up to the current limit $\sim 4 \times 10^{-3}$, Baryogenesis could take place with a branching fraction as low as $2 \times 10^{-4}$. Figure~\ref{fig:BR_fixed} shows that even with a negative  $A_{\ell \ell}^d$,  as expected in the SM,  the baryonic asymmetry can be generated with $\text{Br}(B \to \phi \xi + \text{Baryon}+X) > 2\times 10^{-3}$ provided that $A_{\ell \ell}^s \sim \mathcal{O}(10^{-3})$. We reiterate that both the leptonic asymmetry and the decay of a $B$ meson to a baryon and missing energy are measurable quantities at B-factories and hadron colliders (see Section~\ref{sec:constraints}).

\subsubsection*{The Dark Matter Abundance}
As previously argued, in the absence of additional interactions, our set-up generically tends to overproduce the DM since the leptonic asymmetry is $<5\times10^{-3} $.
By examining the DM yield curve in Figure~\ref{fig:Y_Baryo} we see that annihilations (the dip in the curve) deplete the DM abundance that would otherwise be overproduced from the $\Phi$ decay.

Recall that for a stable particle species annihilating into two particles in the early Universe when neglecting inverse-annihilations: $\Omega h^2 \propto x_{\text{FO}}/\left<\sigma v \right>$. For WIMPs produced through thermal freeze-out $x_{\text{FO}}= m_{\text{DM}} /T_{\text{FO}} \sim 20$, while, in our scenario $m_{\rm DM}/T \sim 400$. 
Therefore, an annihilation cross section roughly 1 order of magnitude higher than that of the usual WIMP is required to obtain the right DM abundance. We have analyzed the extrema of the parameter space and found that we require the dark cross section to be
\begin{align}
\label{eq:DarkCS}
\langle \sigma v \rangle_{\rm dark} &= (20 - 70) \langle \sigma v \rangle_{\rm WIMP} \\ \nonumber
&=\, (6-20) \times 10^{-25} \,\text{cm}^3/\text{s} \,,
\end{align}
where $  \left<\sigma v \right>_{\rm WIMP} = 3\times 10^{-26}\,\text{cm}^3/\text{s}$, and the spread of values correspond to varying the DM mass over the range specified in Table~\ref{tab:Parameters} (with only a very slight sensitivity to other parameters). 
In particular $\left<\sigma v \right>_{\rm dark} \simeq 25 \left<\sigma v \right>_{\rm WIMP}  \, \text{min}[m_{\phi},m_\xi]/\text{GeV} $.

\subsubsection*{Primordial Antimatter with a low Reheat Temperature}

Finally, note that since we are considering rather low reheat temperatures, there could be a significant change to the primordial antimatter abundance. In the case of a high reheat temperature scenario, the primordial antinucleon abundance is tiny: $Y_{\bar{N}} = 10^{18} \times e^{-9\times10^5}$~\cite{Kolb:1990vq}. In our scenario, we can track the antinucleon abundance from the following Boltzmann equation:
\begin{align}
&\dot{n}_{\bar{N}}  + 3 H n_{\bar{N}} = -\left<\sigma v�\right> n_{\bar{N}} n_N +  f_{\bar{N}} \Gamma_\Phi n_\Phi�\nonumber  \\
&= -\left<\sigma v�\right> n_{\bar{N}} \left(n_{\bar{N}} +n_{\phi}-n_{\phi^\star}\right)+  f_{\bar{N}} \Gamma_\Phi n_\Phi 
\end{align}
where $f_{\bar{N}} \simeq 1$ is the produced number of antinucleons per $\Phi$ decay~\cite{Cirelli:2010xx}. By solving this Boltzmann equation we find that for $\Gamma_\Phi > 3\times10^{-23}\,\text{GeV}$ the primordial antinucleon abundance is $Y_{\bar{N}} < 10^{-26}$ (and usually way smaller) and  too small to have any phenomenological impact at the CMB or during BBN.

\bigskip
\section{Searches and Constraints}
\label{sec:constraints}
Developing a testable mechanism of Baryogenesis has always been challenging.  Likewise, should a DM detection occur, nailing down the specific model in set-ups where a rich hidden dark sector is invoked, is generally daunting. 
 The scenario described in the present work is therefore unique in that it is potentially testable by future searches at current and upcoming experiments, while being relatively unconstrained at the moment.

\subsection{Searches at LHCb and Belle-II}
As discussed above, a positive leptonic asymmetry in $B$ meson oscillations -- and the existence of the new decay mode of $B$ mesons into visible hadrons and missing energy -- would both indicate that our mechanism may describe reality. Both these observables are testable at current and upcoming experiments.  
\subsubsection*{Semileptonic Asymmetry in $B$ decays}
As shown in Section~\ref{sec:results}, the model we present requires a positive and relatively large leptonic asymmetry: $A_{\ell \ell} \sim 10^{-5}-10^{-3}$.
The current measurements of the semileptonic asymmetry~\cite{pdg} (recall that in our setup the semileptonic and leptonic asymmetries may be used interchangeably) are: 
\begin{align}
\nonumber
\label{eq:AslMeas}
A_{SL}^s &= (-0.0006\pm 0.0028) \, ,\\ 
A_{SL}^d &= (-0.0021\pm 0.0017) \, .
\end{align}
These are extracted from a combination of various analyses of LHCb and $B$-factories. 
Future and current experiments will improve upon this measurement.
In particular, the future reach of LHCb with $50\,\text{fb}^{-1}$ for the measurement of the leptonic asymmetry is estimated to be $\sigma(A_{\rm 
SL}^s) \sim 5\times 10^{-4}$~\cite{Artuso:2015swg}, and a similar sensitivity should be expected for $A_{SL}^d$. The sensitivity of Belle-II to the semileptonic asymmetries has not been addressed in the Belle-II physics book~\cite{Kou:2018nap}. However, we became aware~\cite{private} that with $50\,\text{ab}^{-1}$ Belle-II should reach a sensitivity to $A_{\rm SL}^d$ of $5\times 10^{-4}$. In addition, Belle-II is planning on collecting $1\,\text{ab}^{-1}$ of data at the $\Upsilon(5S)$ resonance~\cite{Kou:2018nap} which could potentially result in a measurement for $A_{SL}^s$.
\subsubsection*{$B$ meson decays into a Baryon and missing energy}
For our mechanism to produce the observed Baryon asymmetry in the Universe, from Figure~\ref{fig:BR_fixed}, we notice that  moderately large $\text{Br}({B  \to \xi \phi+\text{Baryon}}+ X) = 10^{-4}-0.1$ are required. The constraint on $\text{Br}({B  \to \xi \phi+\text{Baryon}}+ X) < 0.1$ at 95\% CL~\cite{pdg} is based on the measurement of the $B^+$ decay to $cX$. This branching fraction can also be constrained since the presence of this new decay mode could alter the total width of $b$-hadrons. We proceed as~\cite{Brod:2014bfa} and use the theoretical expectation in the SM for the decay width from~\cite{Krinner:2013cja} $\Gamma^b_\text{SM} = (3.6\pm0.8)\times10^{-13}\,\text{GeV}$, and the observed width~\cite{pdg} $\Gamma^b_\text{obs} = (4.202\pm0.001)\times10^{-13}\,\text{GeV}$. This constraint therefore restricts $\text{Br}({B  \to \xi \phi+\text{Baryon}}+ X) < 0.37$ at 95\% CL.

To our knowledge there are no searches available to measure this branching fraction, and no published data on the inclusive branching fraction for $B$ mesons $\text{Br}({B  \to \text{Baryon}}+ X)$ either. We expect that existing data from Babar, Belle and LHC can already be used to place a meaningful limit. The search for this channel should in principle be similar to other $B$ meson missing energy final states such as $B\to K \nu\nu$ or $B\to \gamma \nu\nu$ with current bounds at the level of $\mathcal{O}(10^{-5})$~\cite{pdg}. Given this, the reach of Belle-II~\cite{Kou:2018nap} could be of $\mathcal{O}(10^{-6})$. Thus, potentially    our mechanism is fully testable. 

\newpage
\subsubsection*{Exotic $b$-flavored Baryon decays}
Our Baryogenesis and DM production mechanism requires the presence of the new exotic $B$ meson decays. However, once these decays are kinematically allowed, the $b$-flavored baryons will also decay in an apparently baryon violating way to mesons and DM in the final state. For instance, given the interaction~\eqref{eq:usbOp} the $\Lambda_b^0$ baryon could decay into $ \bar{\psi}+ K^+ + \pi^-$ provided $m_\psi < 4.9\,\text{GeV}$ which will always be the case since $m_\psi < m_{B}-m_\Lambda$ in this case. In addition, the rate of this process should be very similar to that of $B$ mesons.  To our knowledge there is no current search for this decay channel, but in principle LHCb could search for it. In particular, Ref.~\cite{Stone:2014mza} pointed out that it is possible to identify the initial energy of a $\Lambda_b$ if it comes from the decay of $\Sigma_b^\pm, \, \Sigma_b^\pm{}^\star \to \Lambda_b + \pi^\pm$ by measuring the kinematic distribution of the process. The LHCb collaboration has very recently observed $\sim 20000$~\cite{Aaij:2018tnn} $\Lambda_b$ candidates produced via this process, thus making the measurement of $\Lambda_b \to \bar{\psi} + \text{mesons}$ potentially viable at hadron colliders. We refer to Appendix~\ref{sec:decayoperators} where the lightest states of the possible decay processes are outlined for the four different flavor operators.

\subsubsection*{Considerations from Flavor Observables}
Recall that, the semileptonic asymmetry may be computed theoretically from the off-diagonal elements of the Hamiltonian describing the $B^0-\bar{B}^0$ system:
\begin{align}
\label{eq:Hosc}
& \mathcal{H}_{\text{osc}} = 
\left[ \begin{array}{cc}
M_{11} - \frac{i}{2} \Gamma_{11} & M_{12} - \frac{i}{2} \Gamma_{12}   \\
M_{12}^* - \frac{i}{2} \Gamma_{12}^* & M_{22}  - \frac{i}{2} \Gamma_{22}   \end{array} \right]  \, .
\end{align}
In particular,
\begin{equation}
 A^q_{SL}=\text{Im}\left({\frac{\Gamma_{12}^{q}}{M_{12}^q}} \right) \,.
\end{equation}
In the SM, oscillations arise from quark-$W$ box diagrams, whose vertices contain CPV phases from the CKM matrix. 
In particular, the SM predictions for $ A^q_{SL}$ read $A^s_{\ell \ell}|_{\textrm{SM}}  = (2.22\pm 0.27) \times 10^{-5}$ and $A^d_{SL}|_{\textrm{SM}}= (-4.7\pm 0.6)\times 10^{-4}$~\cite{Lenz:2011ti,Artuso:2015swg}. Since these are substantially smaller then the current measurements \eqref{eq:AslMeas} there is room to accommodate new physics. 

The large positive leptonic asymmetry required in our set-up could differ considerably from the SM values, depending on the value of $\text{Br}({B  \to \xi \phi+\text{Baryon} + X})$. 
There are many BSM models that allow for a substantial enlargement of the semileptonic asymmetries of both the $B_d^0$ and $B_s^0$ systems (see \textit{e.g.}~\cite{Artuso:2015swg,Botella:2014qya} and references therein). In addition, it is worth mentioning, that the flavorful models invoked to explain the recent $B$-anomalies also induce sizable mixing in the $B_s^0$ system (see \textit{e.g.}~\cite{Altmannshofer:2014cfa,Celis:2015ara,Gripaios:2014tna,Becirevic:2016yqi}).

Note, that while the elements of the evolution Hamiltonian \eqref{eq:Hosc} are not directly probed in experiment, they can be related to additional experimental observables as:
\begin{align}
\Delta m_{B} = 2 |M_{12}| \, \quad \text{and} \quad \Delta \Gamma_B  = - \frac{2 \text{Re} (M_{12}^* \Gamma_{12})}{|M_{12}|} \, ,
\end{align}
where, for instance $\Delta m_{B}$, the $B$ meosn oscillation length, is related to the mass eigenstates. 
Therefore, any new physics that modifies $A^q_{SL}$ away from the SM value will also modify $\Delta m_{B}$ and $\Delta \Gamma_B$, and must not be in conflict with current bounds on these observables.  For an overview of the allowed BSM modifications to  $\Delta m_{B}$ and $\Delta \Gamma_B$ see \cite{Artuso:2015swg}.

\bigskip
\subsection{Constraints}
Here we comment on collider, cosmological, and DM direct detection constraints.

\subsubsection*{Collider Constraints}
Within our set-up, the heavy colored scalar $Y$ is responsible for inducing the $B$ meson decays into the dark sector via Equation~\eqref{eq:Lag_psi}. The colored scalar may be produced at colliders and its decay products searched for, thus resulting on an indirect constraint on the model. This branching ratio was calculated in~\cite{Aitken:2017wie} and we quote the result here;
\begin{align}
\Gamma_{\bar{b}\to\phi \xi {u} {s}} & \sim\frac{m_b\Delta m^4}{60\left(2\pi\right)^3}\left(\frac{y_{ub} y_{\psi s}}{m_Y^2}\right)^2+{\cal O}\left(\frac{\Delta m^5}{m_b^5}\right)\, ,
\end{align}
from which;
\begin{align}
&\text{Br}({B  \to \xi \phi+\text{Baryon}}) \simeq \\ \nonumber
&10^{-3} \left(\frac{m_B-m_\psi}{2~\rm GeV}\right)^4  \left(\frac{1~\rm TeV}{{m_Y}} \frac{\sqrt{y_{ub} y_{\psi s}}}{0.53} \right)^4 \,.
\label{eq:bdecayrate}
\end{align}

From Equation~\eqref{eq:Lag_psi}, note that $Y$ is dominantly pair-produced at colliders through the strong interaction. The produced $Y$'s could then decay as either $Y \to \bar{u} \, b$ or $Y \to s \, \bar{\psi}$. So that the expected collider signatures are 4-jets (two tagged $b$ quarks) or 2-jets plus missing energy. If the former decay dominates then 4-jet searches~\cite{Aaboud:2017nmi} apply, implying a bound on the colored scalar mass of $m_Y > 500\,\text{GeV}$. While, if $Y \to s \bar{\psi}$ dominates, then $s$-quark searches apply for a single light quark resulting in the bound $m_Y > 960\,\text{GeV}$~\cite{Aaboud:2017nmi}. Such constraints allow for sizable $\text{Br}({B  \to \xi \phi+\text{Baryon}})\sim 10^{-3}$ with moderately large couplings $\sqrt{y_{ub} y_{\psi s}} > 0.25$, and are thus not in tension with our model's prediction of $\text{Br}({B  \to \xi \phi+\text{Baryon}}+ X) = 2\times10^{-4}-0.1$. 

Another possible constraint arises from the single production of $Y$ particles at the LHC via its couplings to light quarks, resulting in di-jet or monojet signatures. These constraints were considered in Ref.~\cite{Aitken:2017wie} and give mass dependent constraints on the $y_{us}$ and $y_{ud}$ couplings which are far from being in tension with the parameters required for our baryogenesis scenario.

Finally, note that the field $\Phi$ is too weakly coupled to be produced at a collider.

\subsubsection*{Cosmological Constraints}
Our mechanism requires a low reheat temperature $T_{RH} \sim \mathcal{O}(10\,\text{MeV})$. The lower bound on the  reheat  temperature comes from the agreement of CMB and BBN observations on the number of relativistic species in the early Universe. The current bound reads $T_{RH} > 4.7 \, \text{MeV}$~\cite{deSalas:2015glj} at 95\% CL which in turn implies that $\Gamma_\Phi < 3 \times 10^{-23} \, \text{GeV} \equiv 45 \, \text{s}^{-1}$ at 95\% CL, where we take $\Gamma_\Phi = 3 H (T_{RH})$. Note that this bound is not expected to be substantially modified by the Planck 2018 final data release since $N_{\rm eff}^\text{2015} > 2.74$~\cite{Ade:2015xua} and $N_{\rm eff}^\text{2018} > 2.70$~\cite{Aghanim:2018eyx} both at 95\% CL. 

\subsubsection*{Dark Matter Direct Detection}
The DM in our scenario could scatter through protons and neutrons, as in the Hylogenesis model~\cite{Davoudiasl:2010am,Davoudiasl:2011fj}, with signatures similar to those in nucleon decay searches (although with somewhat different kinematics). Such searches would test for the presence of interactions that are not needed in a minimal model for Baryogenesis. However, the existing bounds do not constrain either the Hylogenesis model or the mechanism presented here. Within our model, given the interaction with the $b$ quark, the kinematics preclude a direct scattering to $B$ mesons. The scattering to lighter mesons must therefore be accompanied by a penalty due to a weak loop insertion, which makes the expected rate at nucleon decay experiments negligible unless a larger coupling to light quarks exists.
\bigskip
\section{Possible Depletion Mechanisms}
\label{sec:DarkDetails}

As discussed in Sec~\ref{sec:production}, DM is initially over produced. The DM abundance must be depleted sufficiently in order to obtain the measured relic abundance $\Omega_{\text{DM}}h^2 = 0.12$. This requires annihilations of the dark particles with a thermally averaged cross section of order $10^{-25} \,\text{cm}^3/\text{s}$. We will now outline some possibilities for dynamics which can reduce the symmetric DM component. 

\subsection{Annihilations to Sterile Neutrinos}

Right handed neutrinos $N_R$, are massive singlets under the SM symmetries, and as such provide a simple possible depletion mechanism~\cite{Pospelov:2007mp,Escudero:2016ksa} for the DM particles. For instance, for the case where $m_\phi < m_\xi$, we can introduce another dark, heavy, $\mathbb{Z}_2$ odd, Dirac sterile particle $\Psi$ carrying both baryon and lepton numbers. The interaction 
\begin{align}
\mathcal{L} \quad \subset \quad y_N \, \phi \bar{\Psi} N_R +  \text{h.c.} \,,
\end{align}
allows for the DM, $\phi$, to annihilate via $\phi \, \phi^* \rightarrow N  \, N$, thereby depleting the abundance of the symmetric $\phi-\phi^*$ component. 
For the case where $m_\phi > m_\xi$, we could deplete the excess of $\xi$ particles by introducing a $\mathbb{Z}_2$ odd scalar $\Phi^{'}$, charged under lepton number such that the interaction: 
\begin{align}
\mathcal{L} \quad \subset \quad  y_N\, \xi \Phi^{'} N_R  + \text{h.c.} \,
\end{align}
is allowed. The process $\xi \,\xi \rightarrow N \, N$ can annihilate away the overproduced $\xi$ abundance. 

The annihilation cross section to sterile neutrinos is s-wave and as such is subject to strong constraints from the CMB observations as measured by the Planck satellite~\cite{Aghanim:2018eyx,Leane:2018kjk}: $\left<\sigma v \right>_{v=0} \lesssim \left(1-3\right) \left(m_{\rm DM} / \text{GeV} \right) \times 10^{-27}\,\text{cm}^3/\text{s} $, where the range depends upon the annihilation channel (provided the annihilation is not to neutrinos). We note that both $\phi \, \phi^* \rightarrow N \,  N$ and $\xi \,\xi \rightarrow N  \, N$ processes are s-wave but chirality suppressed~\cite{Escudero:2016ksa} \emph{i.e.}  the $s$-wave contribution is suppressed like $(m_N/m_{\rm DM})^2$ as compared with the $p$-wave. In particular, in the limit in which $m_\xi, \, m_\phi \gg m_N$ and in which the mediators are substantially heavier than the DM, the annihilation cross sections go as;
 \begin{align}
\langle \sigma \, v \rangle_{\xi \xi \to N N} &\,=\, y_N^4 \,\frac{m_N^2}{32 \pi  m_{\Phi '}^4 }\left[1  + \frac{2 m_\xi^2}{3 m_N^2} v^2 \right] \, , \\
\langle\sigma \,v\rangle_{\phi^\star \phi \to N N} & \,=\, y_N^4 \, \frac{  m_N^2 }{8 \pi  m_{\Psi '}^4} \left[ 1 + \frac{m_\phi^2}{6 m_N^2} v^2 \right]\,.
\end{align}
So that in this limit the $p$-wave contribution is significantly enhanced, and CMB constraints are substantially ameliorated. Additionally, the decay of sterile neutrinos can be to invisible particles. For instance, if the $N$ fermion is solely mixed with the $\tau$ neutrino (the least constrained scenario~\cite{Alekhin:2015byh}), provided that $m_N < m_\pi$, the decay will entirely be to $3\,\nu$, and CMB constraints are fully evaded.

\newpage
\subsection{Annihilations to Standard Model Neutrinos}
Consider the same set-up as in the previous subsection, but assume the sterile leptons are heavier than the DM. In this case, mixings could generate couplings which would allow the DM particles to potentially annihilate into active neutrinos~\cite{Macias:2015cna,Gonzalez-Macias:2016vxy}. This annihilation will evade CMB constraints and remain elusive to other experiments due to the weak interactions of neutrinos. 

Generically, the coefficients of the mixing operators that generate the annihilation cross section to active neutrinos will be constrained.
However, for very heavy neutrinos the only constraint on the mixing comes from the invisible width of the $Z$ and various low-energy lepton universal processes in the SM. Such measurements bound the mixing to be $|U|^2 < 10^{-2}$~\cite{deGouvea:2015euy} regardless of the heavy lepton mass and specific flavor structure. The annihilation cross section to neutrinos will scale as $\sim y_N^4 |U|^4/(16 \pi m_{\text{\rm med}}^2)$ which for a mixing that saturates the bound requires Yukawa couplings of $y_N \gtrsim 0.6 \sqrt{m_\text{\rm med} /3 \,\text{GeV} }$.

Unfortunately, while SM neutrinos may be produced in this set-up, we do not expect a detectable signal at neutrino detectors given the required annihilation rate~\eqref{eq:DarkCS}~\cite{Beacom:2006tt}.

\subsection{Annihilations within the Dark Sector}
As discussed in Sec.~\ref{sec:Model}, if the depletion of the symmetric part of DM is too efficient, then there will not be enough DM since the stable dark particle masses have to be $m_\phi < m_B-m_p$ for the $B$ decay process to occur, while the observed baryon asymmetry and DM abundances would require $m_\phi = 5.36\times m_p$. This situation, however, could be different if additional baryons are present in the dark sector. 
For instance, we can add another dark scalar particle $\mathcal{A}$ which carries baryon number $1/3$, and mass  $m_{\mathcal{A}} \sim \frac{5}{3} m_p \sim 1.6 \, \text{GeV}$,  (which is allowed by neutron star constraints) and is $\mathbb{Z}_2$ odd with interactions:
\begin{align}
\mathcal{L} \,\,\, \subset \,\,\, \kappa \, \phi \mathcal{A}^3 + \kappa^{\prime} \, \phi  \phi^*  \mathcal{A}   \mathcal{A}^*  \,\,+ \,\,\text{h.c.}\, .
\end{align}
If $\mathcal{A}$ is lighter than $\phi$, the reactions $\phi+ A^*\leftrightarrow {A}+ {A}$ etc. will eventually turn an excess of $\phi$ particles into an excess of $\mathcal{A}$ particles. If we now deplete the symmetric component of the $\mathcal{A}$ abundance we can arrive at the right DM relic abundance. 

In general the same set-up will hold in a situation where the dark sector contains many baryon number charged states. Rich dark sectors offer a variety of possible production mechanisms in the early Universe \cite{Cheung:2010gj, Cheung:2010gk},  that the initial abundance of $\mathcal{A}$ or other dark sector particles need not be fixed in this setup. 
The details of models with rich dark sectors containing baryon number charged particles is beyond the scope of this paper, and we leave this to to future work. 

\newpage
\section{Conclusions}
\label{sec:conclusion}
In this work we have presented a new mechanism for low temperature Baryogenesis in which both the baryon asymmetry and the DM relic abundance arise from the oscillation and subsequent decay of $B$ mesons to visible baryons and dark sector states. 
We have illustrated our set-up with a model that is unconstrained by di-nucleon decay, does not require a high reheat temperature, and would have unique experimental signals -- a positive leptonic asymmetry in $B$ meson decays, the existence of the new decay of $B$ mesons into a baryon and missing energy, and the $b$-flavored baryons decay into mesons and missing energy. These observables are testable at current and upcoming collider experiments.
In summary, the novel features of our mechanism include the following: 

\begin{itemize}[leftmargin=0.5cm,itemsep = 0.0cm]
\item{Low reheat temperature.}
\item{At least one component of DM charged under baryon number.}
\item{Total baryon number of the Universe remains zero.}
\item{Baryon asymmetry directly related to experimental observables.}
\item{Distinctive experimental signals: \begin{enumerate}
\item Positive semileptonic asymmetry in $B$ meson decays.
\item Charged and neutral $B$ meson decays to a baryon and missing energy.
\item Charged and neutral $b$-flavored baryons decays to mesons and missing energy.
\end{enumerate} }
\item{Testable at both current and upcoming experiments.  }
\end{itemize}

The Baryon asymmetry scales fixed by the following product:
\begin{align}
Y_B 
& \,\, \propto \,\, \sum_{q = s, d} A_{\ell \ell}^q  \times \text{Br}(B_q^0 \to \phi \xi + \text{Baryon} + X)\, ,\nonumber
\end{align}
where $A_{\ell \ell}^q$, the leptonic charge asymmetry (which in our set-up is effectively equivalent to the semi-leptonic asymmetry), is an experimental observable which parametrizes the CPV in the $B_s^0$ and $B_q^0$ oscillation systems. 

We have solved the set of coupled Boltzmann equations and mapped out our model parameter space (allowed by kinematics and current constraints) that accommodates a DM relic abundance and baryon asymmetry that agrees with observation. From Figure~\ref{fig:BRAll} and Figure~\ref{fig:BR_fixed} (respectively), we find that our model predicts the following values for the branching ratio and leptonic asymmetry:
\begin{align}
\nonumber
& \text{Br}({B  \to \xi \phi+\text{Baryon}}+ X) = 2\times10^{-4}-10^{-1} \, , \\ \nonumber 
\end{align}
\vspace{-10.5mm}
\begin{align}
\nonumber
&\text{and} \quad  \sum_{q = s, d}  A_{\ell \ell}^q \sim 10^{-5}-10^{-3} > 0  \,,\\ \nonumber 
&\, \quad \quad \quad \quad \quad \Rightarrow  \quad \quad  Y_B \sim 8.7 \times 10^{-11} \,. \nonumber
\end{align} 
These two observables are currently unconstrained but can be measured in future searches at LHCb and Belle-II. In particular, given that current sensitivity for $B\to K \bar{\nu}\nu$ is at the level of $\mathcal{O}(10^{-6})$~\cite{pdg} our scenario could already be tested with current data from B factories, and we expect our mechanism to be fully testable given the reach of Belle-II~\cite{Kou:2018nap}.

We have seen that there is a region in our parameter space where it is possible to solely generate the baryon asymmetry with $A^s_{\ell \ell}|_{\textrm{SM}}  = (2.22\pm 0.27) \times 10^{-5} > 0$ (although new physics effects are required in this case to simultaneously suppress the negative SM value of $A^d_{SL}|_{\textrm{SM}}= (-4.7\pm 0.6)\times 10^{-4}$). However, generically our parameter space predicts values of $A^q_{\ell \ell}$ larger than the SM expectation. Note that flavorful models invoked to explain the recent B-anomalies also induce sizable mixing in the $B_s^0$ system, and therefore a potential link between our mechanism and the B-anomalies would be a very interesting avenue to explore.

We have seen that the baryon-symmetric component of DM will be generically overproduced, and as such we require additional interactions which allow for DM annihilations to deplete the abundance such that we reproduce the observed value. In particular we find that we require the dark sector cross section to be:
\begin{align}
\nonumber
\left<\sigma v \right>_{\rm dark} 
&=\, (6-20) \times 10^{-25} \,\text{cm}^3/\text{s} \,, 
\end{align}
\vspace{-8mm}
\begin{align}
\nonumber
\quad \quad \Rightarrow   \quad  \Omega_{\rm DM} h^2 \sim 0.12 \,.
\end{align}
While, we have preliminarily outlined some classes of models that could explain the depletion of DM, formulating a 
detailed and complete model is one avenue for future investigation. 

Another future direction on the model building front, would be to embed our mechanism in a more detailed UV model to explain the origin and nature of the colored scalar $Y$. Additionally, as discussed above, there is theoretical motivation for multiple dark sector states charged under Baryon number. An interesting investigation to pursue would be to consider various scenarios for dark sector states charged under baryon number.

\acknowledgments

This project has received support from the European Union's Horizon 2020 research and innovation programme under the Marie Sk\l{}odowska-Curie grant agreement No 690575 (InvisiblesPlus RISE) and No 674896 (Elusives ITN). ME is supported by the European Research Council under the European Union's Horizon 2020 program (ERC Grant Agreement No 648680 DARKHORIZONS). ME acknowledges the hospitality of the Particle, Field, and String Theory group at the University of Washington where this work was initiated. GE and AEN are supported in part by the U.S. Department of Energy, under grant number  DE-SC0011637. AEN is also supported in part by the Kenneth K. Young Memorial Endowed Chair. AEN acknowledges the hospitality of the Aspen Center for Physics, which is supported by National Science Foundation grant PHY-1607611. 
We would like to acknowledge David McKeen, Sheldon Stone and Ville Vaskonen for their very useful comments on the first version of this draft. 

\appendix

\section*{Appendices}
Here we give some details behind the calculations in the main text. Appendix~\ref{app:coherence} and~\ref{subsec:BannOrdec} justify the assumptions we made in writing down the Boltzmann equation of Section~\ref{subsec:Boltzmann}. In Appendix~\ref{app:darkcrosssecs} we itemize the dark sector annihilation cross sections. Finally, flavorful variations of the $B$ meson decay operator~\eqref{eq:usbOp} are outlined in Appendix~\ref{sec:decayoperators}.  

\subsection{Decoherence due to Elastic Scattering}
\label{app:coherence}
In our mechanism, the coherence in the $B^0$-$\bar{B}^0$ oscillation system in the early Universe is key to generating the baryon asymmetry. The elastic scattering of $e^\pm B_0 \to e^\pm B_0$ represents the only possible source of decoherence in the early Universe, and in this subsection we calculate the thermal rate of this process in the expanding Universe. 

As the $B_0$ is neutral pseudoscalar particle the only possible interaction that an electron can have with it is through the effective charge distributed within it. This charge distribution is parametrized in terms of a an elastic electromagnetic form factor $F_{B_0}(q^2)$. The actual form of $F_{B_0}(q^2) $ requires either data (which is not possible to obtain in the laboratory for this reaction) or some modeling of the quarks distributed within the $B_0$ meson. The form factors are usually parametrized in terms of the charge radius which is defined as: 
\begin{align}
\langle r^2\rangle = 6 \left[\frac{d F(q^2)}{d q^2} \right]_{q=0} \, ,
\end{align} 
which for a neutral particle leads to form factors:
\begin{align}\label{eq:formfac}
F(q^2) = - \frac{1}{6} \langle r ^2\rangle q^2 + ...\, .
\end{align} 
Since $\langle r_{B_0}^2\rangle$ is not measured, we use an estimate provided by~\cite{Hwang:2001th}; $\langle r_{B_0}^2\rangle  \sim -0.187 \, \text{fm}^2$. For comparison, the measured value for $K_0$ is the following: $ \langle r_{K^0}^2\rangle = - 0.077\pm0.010 \, \text{fm}^2$~\cite{pdg}. 

\begin{figure}
\centering

\includegraphics[width=0.47\textwidth]{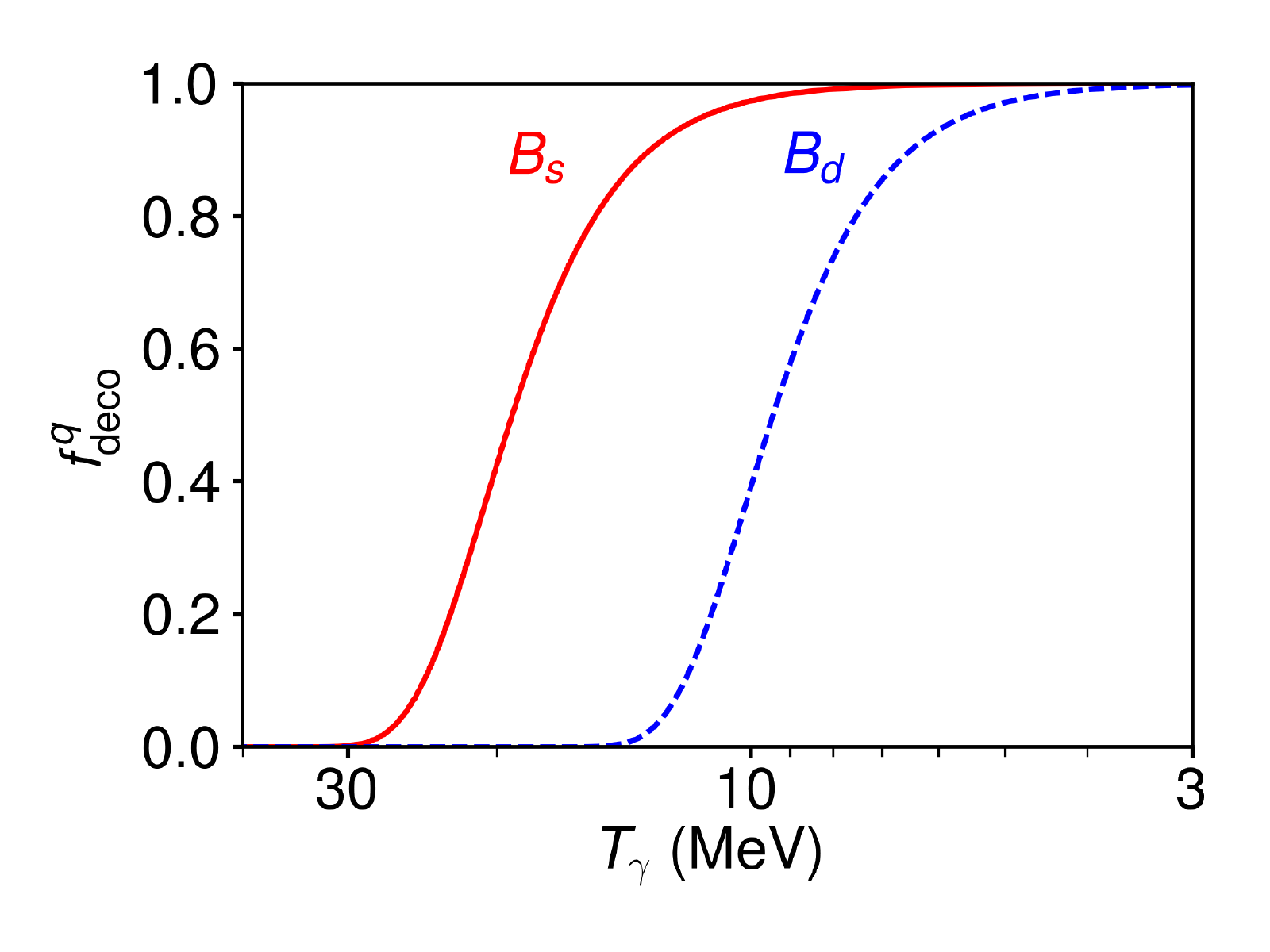}
\vspace{-0.5cm}
\caption{The decoherence function~\eqref{eq:dec_f} as a function of temperature for the $B_s^0$ and $B_d^0$ systems. }
\label{fig:fdeco}
\end{figure}

We can safely use the quadratic expansion for the form factor, \eqref{eq:formfac}, since it will be valid for $|q| < 1/\sqrt{\left< r_{B_0}^2\right> } \sim 100 \, \text{MeV}$ and we are interested in $T \sim 10 \, \text{MeV}$. Since the quadratic form factor approximation holds, we can calculate the differential scattering cross section for the process $e^\pm B_0 \to e^\pm B_0$, which in the lab frame (and ignoring the $B_0$ recoil) reads\footnote{This equation is the non-relativistic formula given for an electron interacting with a target with charge density $\rho$ where $F(q^2) \equiv \int \rho (r) \, e^{i \vec{q} \vec{r}} \, d^3 \vec{r}  $. }:
\begin{eqnarray}
\frac{d\, \sigma}{d \, \Omega} &=& \frac{\alpha^2}{4 E^2 \sin^4 \left( \theta /2 \right)} \cos^2 \left( \theta/2 \right) | F_{B_0}(q^2)|^2 \, ,
\end{eqnarray}
where $\alpha$ is the fine structure constant. The momentum exchanged is:
\begin{eqnarray}
q^2 = -\frac{2 m_{B_0} E^2 (1-\cos \theta)}{ m_{B_0} + E (1-\cos \theta)}  \simeq -4 E^2 \sin^2 \frac{\theta}{2} \, ,
\end{eqnarray}
where $E$ is the energy of the incoming electron. We can therefore rewrite the differential cross section as:
\begin{eqnarray}
\frac{d \, \sigma}{d \, q^2} &=& - \, 2 \, \pi \, \frac{\alpha^2}{18} \, \langle r_{B_0}^2\rangle^2 \,  \left(1+\frac{q^2}{4E^2} \right)\, .
\end{eqnarray}
Upon integration we obtain the total scattering cross section
\begin{eqnarray}
\sigma &=& \int_{-4E^2}^\text{0}\frac{d \, \sigma}{d \, q^2} \, d \, q^2 = \alpha^2 \, \frac{2\pi}{9} \, \langle r_{B_0}^2\rangle^2 \, E^2 \, .
\end{eqnarray}
By substituting the energy $E$ by its average in the early Universe; $E\sim 3T$, we obtain the thermally averaged rate for this process:
\begin{eqnarray}\label{eq:rate_eB-eB}
\Gamma_{e^\pm B_0 \to e^\pm B_0} \,\,  &\equiv&\,\,  \langle \sigma v \rangle n_e \simeq \sigma(E= 3T)  \, n_e(T) \\ \nonumber
&\simeq& \,\, 10^{-11} \, \text{GeV} \left( \frac{T}{20 \, \text{MeV} }\right)^5 \, \left( \frac{\langle r_{B_0}^2\rangle}{0.187 }\right)^2\,.
\end{eqnarray}
Notice that the $e^\pm B_0 \to e^\pm B_0$ scattering rate will be higher than the $B_s$ oscillation rate $\Delta m_{B_{s}} = 1.17 \times 10^{-11}\,\text{GeV}$ for temperatures above $\sim 20\,\text{MeV}$ and therefore through the Zeno effect electron/positron scatterings will damp the $B^0$-$\bar{B}^0$ oscillations. An identical analysis applies for the $B_d$ oscillations, but since $\Delta m_{B_d} = 3.34\times 10^{-13}\,\text{GeV}$ oscillations will only be efficient for $T<10\,\text{MeV}$. We note that the rate calculated in~\eqref{eq:rate_eB-eB} has a very strong temperature dependence and therefore is fairly independent on possible unaccounted details in the $B_0$ form factor. In Figure~\ref{fig:fdeco} we show the resulting decoherence function~\eqref{eq:dec_f} for the $B_s$ and $B_d$ systems given the interaction rate in~\eqref{eq:rate_eB-eB} and the $B$-meson mass differences.

\subsection{Do $B$ Mesons Decay or Annihilate?  }
\label{subsec:BannOrdec}
In the Boltzmann equations of Section~\ref{subsec:Boltzmann}, we have omitted the possibility that $B$ mesons annihilate prior to their decay into the dark sector. In this subsection, we explicitly show that, at the temperatures of interest $T \sim 20\,\text{MeV}$, this is indeed a valid approximation. Additionally, we now determine in which range of temperatures the decays will dominate over annihilations.

Since the $\Phi$ particle decays at the same rate to $B$ and $\bar{B}$, we can assume that $n_B = n_{\bar{B}}$ upon hadronization. In reality, due to CP-violating oscillations, $n_{B} \simeq (1+10^{-3})n_{\bar{B}}$, but this will not impact the calculation at hand. 
The Boltzmann equation that governs the the $B$ number density is:
\begin{align}\label{eq:nBdensity}
 \frac{d n_B}{ dt} + 3 H n_B =  \Gamma_{\Phi} \text{Br}_{\Phi \rightarrow B} n_{\Phi} - \Gamma_B n_B - \langle \sigma v \rangle n_B^2  \, . 
\end{align}
Equation~\eqref{eq:nBdensity} involves very different time scales, and its numerical solution will require time steps of $t < 1/\Gamma_B$ -- which are $10^{-10}$ smaller than those of the $\Phi$ lifetime. 

We determine if a produced $B$ meson will decay or annihilate as follows: integrate Equation~\eqref{eq:nBdensity} with only the first term on the right hand side (so that we ignore both $B$ decay and oscillation), this will give us the maximum number density of $B$'s prior to decay $\Delta n_B$. We can then compare the $B$ decay and the annihilation rates, in order to determine which one dominates.  Integration of the first term in Equation~\eqref{eq:nBdensity} in the time interval $t \to t + 1/\Gamma_B$ leads to:
\begin{align}
\Delta n_B &= \int_t^{t+1/\Gamma_B} \frac{d n_B}{ dt} (t') dt'  \\ \nonumber
&= \int_t^{t+1/\Gamma_B} \Gamma_{\Phi} n_{\Phi}(t') dt' = \frac{\Gamma_{\Phi}}{\Gamma_{B}}    n_{\Phi}(t) \,.
\end{align}
Now, we can clearly compare the decay and annihilation rates:
\begin{align}\label{eq:rates_nb_annordec}
\frac{\Delta n_B \Gamma_B}{\Delta n_B^2 \left< \sigma v \right>} &= \frac{\Gamma_{B}^2}{\Gamma_{\Phi}  \left< \sigma v \right> n_{\Phi}(t) } \,.
\end{align}
When solving numerically for the $\Phi$ number density we found that even with an annihilation cross section of $\langle \sigma v \rangle = 10 \, \text{mb}$, the decay rate overcomes the annihilation rate for $T \lesssim 60\,\text{MeV}$ even for $\Gamma_{\phi} =  10^{-21}\,\text{GeV}$ (and $T \gtrsim 120\,\text{MeV}$ for $\Gamma_{\phi} =  10^{-22}\,\text{GeV}$). Thus, for practical purposes it is safe to ignore the effect of annihilations in the Boltzmann equations.

\subsection{Dark Cross Sections}\label{app:darkcrosssecs}
Here we list the dark sector cross sections to lowest order in velocity $v$ that result from the interaction~\eqref{eq:DarkYukawa}:
\begin{align}
&\sigma_{\phi^\star \phi \to \xi \xi} =  \frac{y_d^4 \left(m_{\xi }+m_{\psi }\right)^2 \left[ \left(m_{\phi }-m_{\xi }\right) \left(m_{\xi }+m_{\phi }\right)\right]^{3/2}}{2 \pi  m_{\phi }^3 \left(-m_{\xi }^2+m_{\psi }^2+m_{\phi }^2\right)^2} \, , \nonumber\\
&\sigma_{\xi \xi \to \phi^\star \phi} |_{m_\phi \to 0} =\frac{v^2 y_d^4}{48 \pi   \left(m_{\xi }^2+m_{\psi }^2\right)^4} \quad �\times  \\
& \quad \quad \quad \quad \quad \quad  \left[2 m_{\xi }^5 m_{\psi }+5 m_{\xi }^4 m_{\psi }^2+8 m_{\xi }^3 m_{\psi }^3 \right. \nonumber \\
&  \quad \quad \quad \quad \quad \quad \quad \quad \left. + \,\, 9 m_{\xi }^2 m_{\psi }^4+6 m_{\xi } m_{\psi }^5+3 m_{\xi }^6+3 m_{\psi }^6\right] \,.\nonumber
\end{align}

\subsection{$B$ Meson Decay Operators}\label{sec:decayoperators}
Here we categorize the lightest final states for all the quark combinations that allow for $B$ mesons to decay into a visible baryon plus DM, and for $\Lambda_b$ baryons decaying to mesons and DM. Note that the mass difference between final and initial states for the $B$-mesons will give an upper bound on the dark Dirac fermion $\psi$ mass. 
In Table~\ref{tab:hadronmasses} we list the minimum hadronic mass states for each operator.
\begin{table}[h]
\label{Table:trans}
\renewcommand{\arraystretch}{1.4}
  \setlength{\arrayrulewidth}{.25mm}
\centering
\small
\setlength{\tabcolsep}{0.18 em}
\begin{tabular}{ |c || c | c | c  |}
    \hline
    Operator &  Initial State &  Final state  				&   $\Delta M$ (MeV)    \\ \hline \hline
\multirow{4}{*}{$\psi \, b\, u\, s$}    &   	$B_d$   &  $\psi + \Lambda \,(usd)$&4163.95\\ 
		  &   	$B_s$    &   $\psi + \Xi^0 \,(uss)$	&4025.03\\ 
  		    & 	$B^+$    &  $\psi + \Sigma^+ \,(uus)$	  	& 4089.95\\
    		   &   	$\Lambda_b$   &  $\bar{\psi} + K^0 $		        &5121.9\\  \hline\hline 
\multirow{4}{*}{$\psi \, b\, u\, d$}   &   	$B_d$   &  $\psi + n \,(udd)$		        &4340.07\\ 
		  &   	$B_s$    &   $\psi + \Lambda \,(uds)$		&4251.21\\ 
 		  &   	$B^+$    &  $\psi + p \,(duu)$	  		& 4341.05			    \\ 
    		   &   	$\Lambda_b$   &  $\bar{\psi} + \pi^0 $		        &5484.5\\  \hline\hline 
\multirow{4}{*}{ $\psi \, b\, c\, s$}  &   	$B_d$   &  $\psi + \Xi_c^0 \,(csd)$		&2807.76\\ 
 		    &   	$B_s$    &   $\psi + \Omega_c \,(css)$	&2671.69\\ 
	           &   	$B^+$    &  $\psi + \Xi^+_c \,(csu)$	  	& 2810.36			    \\
    		   &   	$\Lambda_b$   &  $\bar{\psi} + D^-+ K^+$		        &3256.2\\  \hline\hline
\multirow{4}{*}{$\psi \,b\, c\,d$} &   	$B_d$   &  $\psi + \Lambda_c+ \pi^- \,(cdd)$&2853.60\\ 
  		   &   	$B_s$    &   $\psi + \Xi_c^0 \,(cds)$	&2895.02\\ 
	 	  &   	$B^+$    &  $\psi + \Lambda_c \,(dcu)$	  	& 2992.86			    \\ 
		   &   	$\Lambda_b$   &  $\bar{\psi} + \overline{D}^0 $		        &3754.7\\ \hline
\end{tabular}
\caption{Here we itemize the lightest possible initial and final states for the $B$ decay process to visible and dark sector states resulting from the four possible operators. The diagram in Figure~\ref{fig:example_decay} corresponds to the first line. The mass difference between initial and final visible sector states corresponds to the kinematic upper bound on the mass of the dark sector $\psi$ baryon. }
\label{tab:hadronmasses}
\end{table}

\newpage

\bibliography{DC}

\end{document}